\pdfoutput=1
\documentclass[aps,prd,twocolumn,superscriptaddress,preprintnumbers,floatfix,nofootinbib]{revtex4-2}

\usepackage[colorlinks,urlcolor=NavyBlue,citecolor=NavyBlue,linkcolor=NavyBlue,pdfusetitle]{hyperref}
\usepackage{lineno}
\usepackage{cancel}
\usepackage{graphicx}
\usepackage{natbib}
\usepackage{amsmath}
\usepackage{lmodern}
\usepackage[caption=false]{subfig}
\usepackage{color}
\usepackage{dcolumn}
\usepackage{tensor}
\usepackage{bm}
\usepackage{makecell}
\usepackage[T1]{fontenc}
\usepackage[utf8]{inputenc}
\usepackage{microtype}
\usepackage{etoolbox}
\usepackage{amssymb}
\usepackage{mathrsfs}
\usepackage{accents}
\usepackage[normalem]{ulem}
\usepackage[table,dvipsnames]{xcolor}
\usepackage[all]{hypcap}
\usepackage[inline,shortlabels]{enumitem}
\usepackage{braket}
\usepackage{booktabs}
\usepackage{ulem}
\usepackage{array}
\usepackage{diagbox}
\usepackage{color}
\usepackage{colortbl}
\usepackage{hhline}
\usepackage{multirow}
\usepackage{acronym,xspace}
\usepackage{orcidlink}

\newacro{ANN}[ANN]{artificial neural network}
\newacro{CNN}[CNN]{convolutional neural network}
\newacro{aLIGO}[aLIGO]{advanced LIGO}
\newacro{BBH}[BBH]{binary black hole}
\acrodefplural{BBH}[BBHs]{binary black holes}
\newacro{BH}[BH]{black hole}
\acrodefplural{BH}[BHs]{black holes}
\newacro{BNS}[BNS]{binary neutron star}
\acrodefplural{BNS}[BNS]{binary neutron stars}
\newacro{CBC}[CBC]{compact binary coalescence}
\acrodefplural{CBC}[CBCs]{compact binary coalescences}
\newacro{EM}[EM]{electromagnetic}
\newacro{ETG}[ETG]{event trigger generator}
\newacro{EOS}[EOS]{equation of state}
\newacro{FAP}[FAP]{false alarm probability}
\newacro{FAR}[FAR]{false alarm rate}
\newacro{FPR}[FPR]{false positive rate}
\newacro{GCN}[GCN]{General Coordinates Network}
\newacro{GPR}[GPR]{Gaussian process regression}
\newacro{GRB}[GRB]{gamma-ray burst}
\acrodefplural{GRB}[GRBs]{gamma-ray bursts}
\newacro{GWTC3}[GWTC-3]{}
\newacro{GWTC}[GWTC]{Gravitational-wave Transient Catalog}
\newacro{GW}[GW]{gravitational-wave}
\newacro{IMBH}[IMBH]{intermediate black hole}
\newacro{ISCO}[ISCO]{innermost stable circular orbit}
\newacro{karoo}[KGP]{KarooGP}
\newacro{KNN}[KNN]{KNeighborClassifier}
\newacro{LIGO}[LIGO]{LIGO}
\newacro{LVK}[LVK]{LIGO, Virgo, and KAGRA}
\newacro{MDC}[MDC]{Mock Data Challenge}
\newacro{ML}{machine learning}
\newacro{MMA}[MMA]{multi-messenger astronomy}
\newacro{NSBH}[NSBH]{neutron star and black hole}
\newacro{NR}[NR]{numerical relativity}
\newacro{NS}[NS]{neutron star}
\acrodefplural{NS}[NSs]{neutron stars}
\newacro{O2}[O2]{second observing run}
\newacro{O3}[O3]{third observing run}
\newacro{O4}[O4]{fourth observing run}
\newacro{O5}[O5]{fifth observing run}
\newacro{original}{original}
\newacro{predict}{predict}
\newacro{predicted}{predicted}
\newacro{PSD}[PSD]{power spectral density}
\newacro{RF}[RF]{random forest}
\newacro{ROC}[ROC]{Receiver Operating Characteristic}
\newacro{SNR}[SNR]{signal-to-noise ratio}
\newacro{SQZ}[SQZ]{squeezer}
\newacro{TT}[TT]{training+testing}
\newacro{TPR}[TPR]{true positive rate}
\newacro{LVC}[LVC]{LIGO and Virgo Collaboration}

\newcommand{\hasns}{{{\tt HasNS}}}
\newcommand{\hasrem}{{{\tt HasRemnant}}}
\newcommand{\hasgap}{{{\tt HasMassGap}}}

\newcommand{\true}{{{\tt TRUE}}}
\newcommand{\false}{{{\tt FALSE}}}

\graphicspath{{./}{figs/}}

\newcommand{\MST}{\affiliation{Institute of Multi-messenger Astrophysics and Cosmology \& Physics Department, 
		Missouri University of Science and Technology, Rolla, MO 65409, USA}}
\newcommand{\OleMiss}{\affiliation{Department of Physics and Astronomy,
		University of Mississippi, University, Mississippi 38677, USA}}
\newcommand{\Torino}{\affiliation{Dipartimento di Fisica,
		Universit\`{a} di Torino \& INFN Sezione di Torino, via P. Giuria 1, 10125 Torino, Italy}}
\newcommand{\Tubingen}{\affiliation{Theoretical Astrophysics Department, 
		Eberhard-Karls University of T\"{u}bingen, T\"{u}bingen 72076, Germany}}
\newcommand{\UAB}{\affiliation{Departament de Matem\`{a}tiques,
		Universitat Aut\`{o}noma de Barcelona, 08193 Bellaterra, Spain}}
\newcommand{\UV}{\affiliation{Departament d'Astronomia i Astrof\'{i}sica,
		Universitat de Val\`{e}ncia, Dr. Moliner 50, 46100, Burjassot (Val\`{e}ncia), Spain}}
\newcommand{\MNU}{\affiliation{School of Physics and Astronomy, University of Minnesota, Minneapolis, MN 55455, USA}}

\begin{document}

\title{Bayesian real-time classification of multi-messenger electromagnetic and gravitational-wave observations}

\author{Marina Berbel \orcidlink{0000-0001-6345-1798}}
\email{mberbel@mat.uab.cat}
\UAB

\author{Miquel Miravet-Ten\'{e}s \orcidlink{0000-0002-8766-1156}}
\email{miquel.miravet@uv.es}
\UV

\author{Sushant Sharma Chaudhary \orcidlink{0000-0002-5432-1331}}
\email{sscwrk@umsystem.edu}
\MST

\author{\\ Simone Albanesi \orcidlink{0000-0001-7345-4415}}
\Torino
\author{Marco Cavagli\`{a} \orcidlink{0000-0002-3835-6729}}
\MST
\author{Lorena \surname{Maga\~na Zertuche} \orcidlink{0000-0003-1888-9904}}
\OleMiss
\author{Dimitra Tseneklidou \orcidlink{0000-0003-2582-1705}}
\UV
\Tubingen
\author{Yanyan Zheng \orcidlink{0000-0002-5432-1331}}
\MST

\author{\\ Michael W. Coughlin \orcidlink{0000-0002-8262-2924}}
\MNU

\author{Andrew Toivonen \orcidlink{0009-0008-9546-2035}}
\MNU

\hypersetup{pdfauthor={LastName et al.}}

\date{\today}

\begin{abstract}

Because of the electromagnetic radiation produced during the merger, compact binary coalescences with neutron stars may result in multi-messenger observations. In order to follow up on
the gravitational-wave signal with electromagnetic telescopes, it is critical to promptly identify the properties of these sources.  This identification must rely on the properties of the
progenitor source, such as the component masses and spins, as determined by low-latency detection pipelines in real time. The output of these pipelines, however, might be biased, which
could decrease the accuracy of parameter recovery. Machine learning algorithms are used to correct this bias. In this work, we revisit this problem and discuss two new implementations of
supervised machine learning algorithms, K-Nearest Neighbors and Random Forest, which are able to predict the presence of a neutron star and post-merger matter remnant in low-latency
compact binary coalescence searches across different search pipelines and data sets. Additionally, we present a novel approach for calculating the Bayesian probabilities for these two
metrics. Instead of metric scores derived from binary machine learning classifiers, our scheme is designed to provide the astronomy community well-defined probabilities. This would
deliver a more direct and easily interpretable product to assist electromagnetic telescopes in deciding whether to follow up on gravitational-wave events in real time.

\end{abstract}

\maketitle


\section{Introduction} \label{intro}

The first detection of a \ac{GW} signal from a pair of coalescing black holes in 2015 and the first observation of a coalescing binary neutron star system two years later have established
\ac{MMA} as a powerful tool for the exploration of the cosmos~\cite{LIGOScientific:2016aoc,LIGOScientific:2017vwq,LIGOScientific:2018cki}. The third \ac{LVK} catalog of transient \ac{GW}
signals \ac{GWTC3}~\cite{LIGOScientific:2021djp} has shown that \ac{GW} astronomy has entered its mature phase, becoming a true observational branch of astronomy. \ac{MMA} allows
scientists to explore in depth the origin and structure of \acp{NS}~\cite{Ruiz:2021gsv,Baiotti:2016qnr,Lasky:2015uia},
\acp{GRB}~\cite{Murase:2019tjj,Ciolfi:2018tal}, and \acp{BH} through the observation of their progenitors~\cite{Schmidt:2020ekt,Nitz:2021zwj,Barack:2018yly}, test general relativity~\cite{LIGOScientific:2021sio,Berti:2018cxi,Berti:2018vdi,Isi:2019aib}, probe the fundamental nature of
gravity~\cite{LISA:2022yao,Barausse:2020rsu,Piorkowska-Kurpas:2022idw}, and measure the evolution of the universe~\cite{LIGOScientific:2021psn,LIGOScientific:2021aug}.

A wealth of new detections is being amassed to achieve these science goals~\cite{LIGOScientific:2014pky,VIRGO:2014yos}. The rate of detections in \ac{LVK}'s \ac{O4} is close to one per
day and is expected to further increase in the \ac{O5}~\cite{KAGRA:2013rdx}. Over the course of \ac{O4} and the next observing runs, the LVK collaborations are expected to analyze hundreds of binary black hole (BBH) candidate detections, as well as dozens of binary neutron star (BNS) and neutron star and black hole (NSBH) merger events that could potentially be MMA sources. Among the challenges that this new phase of \ac{GW} astronomy brings is the necessity
to coordinate the activities of \ac{EM} and \ac{GW} observatories in real time. 

One of the most interesting areas of study in \ac{MMA} is the physics of gravity-matter interaction in \ac{GW} sources. Tidally disrupted matter in systems with a \ac{NS} may form a
high-temperature accretion disk around the \ac{BH} and trigger the creation of a prompt \ac{EM} emission in the form of a short \ac{GRB}. If the system ejecta are unbound, r-process
nucleosynthesis may lead to a kilonova~\cite{Lattimer:1974slx, Li:1998bw, Korobkin:2012uy, Barnes:2013wka, Tanaka:2013ana, Kasen:2014toa}. These phenomena could also arise in \ac{BNS}
post-mergers through the expulsion of neutron-rich material even when tidal forces are weak~\cite{LIGOScientific:2017ync, Arcavi:2017xiz, Coulter:2017wya, Kasliwal:2017ngb,
Lipunov:2017dwd, DES:2017kbs, Tanvir:2017pws}. The presence of a post-merger matter remnant, which results in an \ac{EM} signature or a prompt collapse, is a common factor in all of these
scenarios. Determining the potential of a \ac{GW} source to become an \ac{EM} emitter and enabling coincident observations of these systems by \ac{EM} and \ac{GW} observatories in low
latency are crucial for the success of \ac{MMA}.

The \ac{LVK} employs different matched-filtering pipelines for low-latency \ac{GW}
searches~\cite{Sachdev:2019vvd,PhysRevD.95.042001,Sachdev:2020lfd,Nitz:2018rgo,Adams:2015ulm,Chu:2020pjv,Klimenko:2015ypf}. These searches are based on discrete template banks of \ac{CBC}
waveforms that provide, among other parameters, the component masses and the dimensionless (anti-)aligned spins of the objects along the orbital angular momentum. These parameters can be
used to determine the \ac{EM} properties of \ac{GW} candidates in real time through empirical fits of \ac{NR} simulations~\cite{Foucart:2012nc,Foucart:2018rjc}. Alerts of candidate
\ac{GW} events that included two \ac{EM}-property metrics identifying whether the \ac{CBC} system contains a \ac{NS}, \hasns, and a post-merger matter remnant, \hasrem, were issued in
\ac{O3} with a median latency of the order of a few minutes after detection. Alerts with similar content continue to be issued in \ac{O4} with even lower latency.  Additionally,
\ac{LVK}'s \ac{O4} alerts include a measure of the \hasgap\ property, i.e., the likelihood that one of the source compact objects has a mass in the lower-mass gap region between \acp{NS}
and \acp{BH}~\cite{Farah:2021qom} (see the \ac{LVK} online user guide~\cite{UserGuide}). 

Classification of \ac{GW} candidate events in low latency poses several challenges as the need for accuracy contrasts with the urgency to issue the information as quickly as possible. The
approach taken by the \ac{LVK} so far has been to use a specific implementation of a supervised \ac{KNN} \ac{ML}
algorithm~\cite{Pedregosa:2011ork} with input from the detection pipelines and \ac{EOS} models to generate independent \hasns\ and \hasrem\ \emph{binary} classification
probabilities~\cite{Chatterjee:2019avs}. The \ac{KNN} model is trained on a broad set of synthetic \ac{CBC} signals injected in real detector noise from the \ac{O2}. The advantage of this
scheme relies on its capability to handle the statistical systematic uncertainties in the parameters recovered by the search pipelines. This approach allowed for a marked improvement in
speed compared to the semi-analytic effective Fisher formalism method that was deployed in \ac{O2}.

In this work, we revisit the problem of the real-time production of \hasns\ and \hasrem\ probability metrics with the aim of further improving the latency and performance of the
current \ac{LVK} \ac{ML}-based scheme~\cite{Chatterjee:2019avs}.  In the current \ac{LVK} implementation, the validity of an event's classification outcome is evaluated using elements from the algorithm's confusion matrix and its \ac{ROC} curve, which is typically used to evaluate the performance of an algorithm by plotting the \ac{TPR} against the \ac{FPR} at various threshold values. \emph{Algorithmic} probabilities for \hasns\ and \hasrem\ are the fractions of \ac{KNN} neighbors around an event that, according to the algorithm's training, indicate whether the event  has an \ac{NS} or is \ac{EM}-bright. In this work, we adopt a new strategy, constructing \emph{Bayesian} probabilities from a complementary data set to the training data set. A second limitation of the current  classification implementation is the fact that the \hasns\ and \hasrem\ labels are treated as independent variables. In reality, the probability of the formation of
tidally disrupted matter, i.e. an \ac{EM} bright event, is always smaller than the probability of the system hosting a \ac{NS}. Therefore, the \hasns\ and \hasrem\ labels cannot be
treated as disjoint. Currently, this physical requirement is not implemented mathematically but rather is based on data; the lack of inversions is tested a posteriori using a technique known as \emph{parameter sweep}, which involves assessing the algorithm's performance across the whole parameter space. One of the main purposes of this work is to go beyond the above scheme and
calculate true \emph{conditional} probabilities for \hasns\ and \hasrem.

To achieve this, we design a new classification scheme and perform a thorough study and comparison of two \ac{ML} algorithms. We first implement a definition for conditional \hasns\ and
\hasrem\ metrics that incorporates \emph{ab initio} the physical requirement that a system with a post-merger matter remnant must necessarily contain a \ac{NS}. Then we calculate Bayesian probabilities for \hasns\ and \hasrem. We also implement a marginalization procedure over a set of \ac{EOS} that minimizes possible systematics arising from the use of a
single \ac{EOS}. Finally, we test the performance of the scheme and the algorithms on synthetic \ac{O3} signals and confident \ac{GW} detections from the latest \ac{GWTC3} catalog.

The structure of the paper is as follows. Section~\ref{algos} introduces the classification algorithms. Bayesian probabilities for \hasns\ and \hasrem\ and the labeling scheme are defined
in Sec.~\ref{probability}. The data set is described in Sec.~\ref{dataset}. Results and algorithm comparisons are reported in Sec.~\ref{results}. Conclusions and future developments are
presented in Sec.~\ref{conclusions}. The process of cross-validation of our algorithms is covered in depth in Appendix~\ref{app:crossval}. A direct comparison of our approach with the current LVK method is provided in Appendix~\ref{app:comparison}.

\section{Classification Algorithms} \label{algos}

We consider two alternative algorithms for unsupervised classification: \ac{KNN} and \ac{RF}. These methods were chosen because of their versatility, ease of implementation in
low-latency, and comparison with the algorithms currently being used by the \ac{LVK} (\ac{KNN} for \hasns\ and \hasrem, \ac{RF} for \hasgap). In this section, we briefly describe the
relevant features of the algorithms and their implementation. 

\subsection*{K-Nearest Neighbors}

K-Nearest Neighbors is a non-parametric, supervised algorithm~\cite{Fix:1951,Cover:1967} that uses the fact that similar points in a data set are ``near'' each other in their parameter
space. When it is applied to classification problems~\cite{Guo:2004}, the algorithm is usually renamed \ac{KNN}. The algorithm captures the idea of similarity between points by computing
the distance between each point in a training set and its neighbors according to a pre-determined metric. Next, it sorts the neighbors in ascending order based on their distance to the
testing point. By choosing the top $K$ neighbors from the sorted array, \ac{KNN} assigns the label to the testing point that corresponds to the most frequent neighbor.

In this work, we use the open-source Python \ac{KNN} implementation of scikit-learn~\cite{Pedregosa:2011ork}. We fix the algorithm hyperparameters by cross-validating over the data set
and obtain the highest accuracy. Throughout our analysis we use $K = 8$ neighbors, the Manhattan metric, the BallTree algorithm and the neighbors are weighted by the inverse of their
distance to the event. Further details on hyperparameter tuning are given in Appendix~\ref{app:crossval}. This configuration differs from the \ac{LVK}'s current implementation, which employs the Mahalanobis metric and $K = 2n + 1 = 11$ neighbors, where $n$ is the number
of features. Our configuration is the optimal choice for the new labeling scheme that is presented in Sect.~\ref{labelingscheme}).

\subsection*{Random Forest}

\ac{RF} is a classification method based on an ensemble of decision trees. The trees are hierarchical models that make decisions by recursively splitting the data at the separation nodes
into different categories based on the values of features. Each tree in the forest is trained independently. To classify a data point, each tree predicts a category and the algorithm
assigns the one that has been chosen the most. \ac{RF} is known for its parallelization capabilities, as computations inside each tree are independent from the rest. An \ac{RF} algorithm
is usually trained using bootstrapping, a technique that randomly assigns subsets of the training data set to each tree. This helps prevent overfitting, as each individual classifier is
not exposed to the same data, and encourages pattern recognition by studying the same data from different subsets. The model's performance on the given data set can be optimized by tuning
the input hyperparameters.

In this work, we use the open-source Python \ac{RF} implementation of scikit-learn~\cite{Pedregosa:2011ork}. The main tunable hyperparameters are the number of trees, the maximum allowed
depth, and the information gain criteria used at splitting. Similar to \ac{KNN}, we choose the algorithm's optimal hyperparameters by measuring the accuracy in the testing set. We use 50
trees in the forest, with a maximum depth of 15, and the maximum number of features considered in a node is the square root of the total number of features. Appendix~\ref{app:crossval} provides details on the cross-validation method used to determine these hyperparameters.

\section{Probability and Labeling Scheme} \label{probability}

Bayesian probabilities for the \hasns\ and \hasrem\ metrics can be defined using the results of the \ac{KNN} and \ac{RF} algorithms, namely the fraction of \ac{KNN}'s neighbors and
\ac{RF}'s trees that predict a given label. In this section, we define the probabilities and explain how to derive them from a data set of simulated signals.  

\subsection{Definition of Bayesian Probabilities} \label{bayesian_probs}

Let us define the probability of a candidate event $E$ being originated by a system with a \ac{NS} [$E(\hasns)=\true$] and \ac{EM} bright [$E(\hasrem)=\true$] given the classifier's outcome evaluated on the detection pipeline
output, ${\bf A}_{\bf X}$, as $P(\hasns|{\bf A}_{\bf X})$ and $P(\hasrem|{\bf A}_{\bf X})$, respectively. The classifier outcome can be understood as a map $A:{\bf X}\to {\bf A}_{\bf X}$,
where ${\bf X}(E)$ is a vector that identifies the output of the detection pipeline and ${\bf A}_{\bf X}$ is a vector that uniquely identifies the classifier algorithm's output for $\bf X$. 

Since a system can be \ac{EM} bright only when a \ac{NS} is present in the system, the condition $P(\hasrem|{\bf A}_{\bf X})\le P(\hasns|{\bf A}_{\bf X})$ must hold. However, if the
probabilities are calculated disjointly, this condition may be violated because of statistical and systematic errors in the pipeline's reconstructed signal parameters, as well as bias and
limited accuracy of the \ac{ML} algorithm. The approach discussed below avoids the occurrence of this inconsistency.

The true properties (ground truth) of an observed event are unknown. Therefore, the probabilities $P(\hasns|{\bf A}_{\bf X})$ and $P(\hasrem|{\bf A}_{\bf X})$ cannot be calculated from observations. However,
estimators for $P(\hasns|{\bf A}_{\bf X})$ and $P(\hasrem|{\bf A}_{\bf X})$ can be calculated from synthetic events under the assumption that these events are a faithful representation of
real observations. This can be done as follows.

According to Bayes' theorem, $P(\hasns|{\bf A}_{\bf X})$  can be written as 
\begin{equation}
P(\hasns|{\bf A}_{\bf X})=\frac{P({\bf A}_{\bf X}|\hasns)P(\hasns)}{P({\bf A}_{\bf X})}\,,
\label{bayes}
\end{equation}
where $P({\bf A}_{\bf X}|\hasns)$ is the likelihood of observing the classifier's outcome given an event with a \ac{NS} in the system, $P(\hasns)$ is the probability that a system includes
a \ac{NS}, and $P({\bf A}_{\bf X})$ is the probability of observing the classifier outcome ${\bf A}_{\bf X}$. Now consider a data set of synthetic events $E'$ defined as
$D=\{{\bf X}(E')\otimes{\bf L}(E')\}$, where ${\bf L}$ is a map to a vector space that assigns \hasns\ and \hasrem\ \emph{labels} given the event's \emph{properties}. We assume that the synthetic set
is a faithful representation of the space of possible real events, i.e., $E \simeq E'$. The probability $P(\hasns|A_{\bf X})$ in Eq.~\eqref{bayes} can be approximated as: 
\begin{align}
\label{bayes-hasns}
P(\hasns|{\bf A}_{\bf X})&\simeq P(\hasns_+|{\bf A}_{{\bf X}'}) \nonumber \\
&=\frac{P({\bf A}_{{\bf X}'}|\hasns_+)P(\hasns_+)}{P({\bf A}_{{\bf X}'})},
\end{align}
where $\hasns_+=\{E'\,|\,E'(\hasns)=\true\}$ identifies the elements in $D$ with positive (+) \hasns\ true property, i.e., the subset of synthetic events that have
been simulated to contain a \ac{NS}, and ${\bf A}_{{\bf X}'}$ is the outcome of the classifier on ${\bf X}' = {\bf X}(E')$. The label $\hasns_+$ is determined by the values of the simulated
parameters of the events $E'$ before they are injected into the detection pipeline. Therefore, the classification label does not depend on the pipeline's outcome. The probability $P(\hasrem|{\bf A}_{\bf X})$ can be approximated as:
\begin{align}
\label{bayes-hasrem}
P(&\hasrem|{\bf A}_{\bf X}) \nonumber  \\
&=P(\hasrem|\hasns, {\bf A}_{\bf X})P(\hasns|{\bf A}_{\bf X}) \nonumber \\
&\simeq P(\hasrem_+|{\bf A}_{{\bf X}''})P(\hasns_+|{\bf A}_{{\bf X}'}),
\end{align}
where $\hasrem_+=\{E'\,|\,E'(\hasrem)=\true\}$ identifies the elements in $D$ with positive (+) \hasrem\ true property, i.e., the subset of synthetic events that have
been simulated to be \ac{EM}-bright, and ${\bf A}_{{\bf X}''}$ is the algorithm's outcome on the subset of events with property $E'(\hasns)=\true$,  ${\bf X}''= {\bf X}'(\hasns_+)$.

To evaluate Eqs.~\eqref{bayes-hasns} and~\eqref{bayes-hasrem} with an \ac{ML} classifier, the synthetic data set is divided into two subsets, $D=D_R\oplus D_S$, where the $\oplus$ sign indicates complementary  subsets. The $D_R$ subset is used
for algorithm training and validation. The $D_S$ subset is used to estimate the probabilities. Throughout this paper we use a 70\% -- 30\% split for $D_R$
and $D_S$, respectively~\cite{split}.

The choice of the labeling scheme and the algorithm's outcome depend on the \ac{ML} algorithm characteristics. Throughout this paper we implement a multi-label classification scheme,
where each element in $D$ is classified into $n$ mutually exclusive categories that uniquely define the $n$ possible physical states of the system. Given that we want to specify two probabilities for \hasns\ and \hasrem, we select the classifier outcome as a vector in a two-dimensional slice of the vector space ${\bf A}_{{\bf X}'}\subset \mathbb{R}^n$. This scheme allows the calculation of Eqs.~\eqref{bayes-hasns} and \eqref{bayes-hasrem} with a single
training process. In the problem at hand, there are three possible physical states. A suitable labeling is

\begin{align}
&{\bf L}[E'(\hasns)=\false]=0\,\nonumber\\
&{\bf L}[E'(\hasns)=\true,\,(\hasrem)=\false]=1\,\\
&{\bf L}[E'(\hasns)=\true,\,(\hasrem)=\true]=2\,.\nonumber
\end{align}

With the above definitions, $\hasrem_+$ is the set of events labeled ``2'' and $\hasns_+$ is the union of the sets labeled ``1'' and ``2''. Therefore a natural choice for the algorithm outcome is
\begin{equation}
{\bf A}_{\bf X'}=(f_1+f_2,f_2)\subset (f_0,f_1,f_2)\,,
\end{equation}
where $f_0({\bf X}')=1-f_1({\bf X}')-f_2({\bf X}')$, $f_1({\bf X}')$, and $f_2({\bf X}')$ are the fractions of \ac{KNN} neighbors or \ac{RF} trees that predict the event to
have labels 0, 1, and 2, respectively.

The factors on the right-hand side of Eqs.~\eqref{bayes-hasns} and~\eqref{bayes-hasrem} can be obtained from $D_{S}$ once the algorithm has been trained on $D_{R}$. For example, for the \ac{KNN} and \ac{RF} scheme described earlier, the factors in Eq.~\eqref{bayes-hasns} can be estimated as  
\begin{align}
&P(\hasns_+)=\frac{N_{\hasns_+}}{N_s}\,,\nonumber\\
&P({\bf A}_{{\bf X}'}|\hasns_+)=\frac{N^+{}_{\hasns_+}(f_1+f_2)}{N_{\hasns_+}}\,,\\
&P({\bf A}_{{\bf X}'})=\frac{N^+{}_{\hasns_+}(f_1+f_2)+N^-{}_{\hasns_+}(f_1+f_2)}{N_s}\,,\nonumber
\label{bayes-est}
\end{align}
where $N_s$ is the number of events in $D_S$, $N_{\hasns_+}$ is the number of  $\hasns_+$ events in $D_S$, and $N^+{}_{\hasns_+}(f_1+f_2)$ and $N^-{}_{\hasns_+}(f_1+f_2)$ are the number of ${\hasns_+}$
events in $D_S$ that are correctly and incorrectly classified by the outcome $f_1+f_2$, respectively. The first factor on the right-hand side of Eq.~\eqref{bayes-hasrem}
can be evaluated similarly to Eq.~\eqref{bayes-hasns} by replacing \hasns\ with \hasrem\ and restricting $D_S$ to \hasns$_+$ elements.

The probability estimators are generally noisy because they are evaluated on a finite data set. Smooth probability functions can be obtained by mapping them from the $(0,1)$ space to the
real line with a logistic function, smoothing them with a Savitzky-Golay filter, fitting them with \ac{GPR}, and finally mapping them back to the $(0,1)$ space.   

Both $P(\hasns|{\bf A}_{\bf X})$ and $P(\hasrem|{\bf A}_{\bf X})$ depend on the \ac{EOS} that is used to label the synthetic events. As the true \ac{EOS} of matter at \ac{NS} densities is
unknown, in order to minimize the systematics that arise in adopting a specific \ac{EOS} we consider a set of 23 different \ac{EOS} and marginalize Eqs.~\eqref{bayes-hasns} and
\eqref{bayes-hasrem} over them. The marginalized probabilities $P_M(\hasns|{\bf A}_{\bf X})$ and $P_M(\hasrem|{\bf A}_{\bf X})$ are defined as
\begin{equation}
\begin{aligned}
P_M(I|{\bf A}_{\bf X})=\frac{\sum_J\beta_J P_J(I|{\bf A}_{\bf X})}{\sum_J\beta_J}\,,
\label{bayes-marginalized}
\end{aligned}
\end{equation}
where $I=$ \hasns\ or \hasrem, $P_J(I|{\bf A}_{\bf X})$ ($J=1,\dots 23$) are the Bayesian probabilities in Eqs.~\eqref{bayes-hasns} and \eqref{bayes-hasrem} calculated from the data set
$D_S$ with labels assigned according to the $J$-th \ac{EOS}, and $\beta_J$ are Bayes' factors from Table~II (third column) of Ref.~\cite{Ghosh:2021eqv}. The probabilities $P_J(I|{\bf A}_{\bf X})$ in
Eq.~\eqref{bayes-marginalized} can be tabulated and used to compute the marginalized probabilities $P_M(\hasns|{\bf A}_{\bf E})$ and $P_M(\hasrem|{\bf A}_{\bf E})$ for any new event $\bf
E$ with algorithm outcome $\bf A_{E}$. It is important to point out that this method does not depend on the specific value of the Bayes' factors.

\subsection{Data Set} \label{dataset}

We use a large data set $D$ of simulated \ac{BNS}, \ac{NSBH}, and \ac{BBH} events that was first used for the space-time volume sensitivity analysis of the \ac{LVK} GstLAL search
\cite{Sachdev:2019vvd,PhysRevD.95.042001,Sachdev:2020lfd} and later employed in Ref.~\cite{Chatterjee:2019avs}. This allows us to directly compare the performance of the various
algorithms and the new labeling scheme to the performance of the \ac{KNN} algorithm that is deployed in the current \ac{LVK} observing run. 

The simulated signals are coherently injected in two-detector data from the \ac{O2} \ac{LVC} observing run. The injection population is built with uniform/loguniform distribution of
component masses whereas the component spins are aligned and injected according to isotropic distributions. Further details on the waveforms and injection parameters can be found in
Ref.~\cite{Chatterjee:2019avs}. The data set $D$ includes approximately 200,000 injected signals that are recovered by the GstLAL pipeline with a \ac{FAR} $\le$ 1/month. The \ac{RF} and
\ac{KNN} algorithms are trained and tested on the injected and recovered intrinsic source properties (primary and secondary masses and spins) and on the recovered \ac{SNR}.

\subsection{Labeling Scheme}\label{labelingscheme}

To label the synthetic data set $D=D_R\oplus D_S$, we follow the practice in use in the \ac{LVK} and identify an event with \hasns:\true\ when at least one of the injected component
masses is less than the maximum \ac{NS} mass allowed by the \ac{EOS}. The value of the maximum \ac{NS} mass ranges from $1.922$ $M_{\odot}$ to $2.753$ $M_{\odot}$ across the various
\ac{EOS}. The lowest and largest maximum \ac{NS} masses correspond to the {\tt BHF\_BBB2} and {\tt MS1\_PP} \ac{EOS}, respectively. We will highlight these two cases together with the
{\tt SLy} \ac{EOS}, which is the most accepted \ac{EOS} in the astrophysics community. We set the \hasrem\ label as \hasrem:\true\ and \hasrem:\false\ for \ac{BNS} and \ac{BBH} systems,
respectively. The value of the \hasrem\ label for \ac{NSBH} events depends on the \ac{EOS} of the \ac{NS}. To identify the \hasrem\ event class for \ac{NSBH} systems,  we follow
Ref.~\cite{Chatterjee:2019avs} and apply Eq.~(4) from Ref.~\cite{Foucart:2018rjc}, colloquially known as the \emph{Foucart formula}.

The Foucart formula is an empirical fit that predicts the total mass of the accretion disk, the tidal tail, and the ejected mass from the final \ac{BH}. The main parameters of the fit are
the compactness of the \ac{NS},  the \ac{NSBH} binary system's symmetric mass ratio, and the normalized \ac{ISCO} radius. The tidal disruption of the \ac{NS} is affected by the mass and
spin of the \ac{BH}. A highly spinning, low-mass \ac{BH}'s small ISCO allows the \ac{NS} to inspiral closer to the \ac{BH} and its tidal force to tear the \ac{NS} apart, resulting in
matter ejecta. If the tidal forces are weak or the \ac{NS} is very compact, the \ac{BH} will swallow the \ac{NS} and there will be no remnant mass.

Different \ac{EOS}s have different thresholds for the mass of the remnant. We label events with inferred masses less than this threshold as \hasrem:\true. For events with component masses
between $\sim 2.5$ and $3.5 M_{\odot}$, the stiffness of the \ac{EOS} is the main factor determining the \hasrem\ label.

\section{Results} \label{results}

In this section, we first discuss the performance of the trained \ac{RF} and \ac{KNN} classifiers on the testing data set $D_S$ and use the latter to calculate the Bayesian probabilities
$P_M(I|{\bf A})$. Then we evaluate $P_M(I|{\bf A})$ on two independent data sets. The first set includes a population of simulated \ac{CBC} events that were injected in the real-time
replay of \ac{O3} data and was used for the \ac{LVK} \ac{MDC}~\cite{Chaudhary:2023vec}. The second set contains the confident \ac{LVK} \ac{O3} detections that are reported in \ac{LVK}'s \ac{GWTC}~\cite{2023ApJS..267...29A}.

\subsection{Performance on the O2 Testing Set}

We assess the performance of the classifiers by measuring the \ac{TPR} and the \ac{FPR} of the events in $D_S$. We present our findings as \ac{ROC} curves that illustrate how the \ac{TPR}
varies for various score thresholds as a function of the \ac{FPR}.

\begin{figure*}
\includegraphics[width=0.45\linewidth]{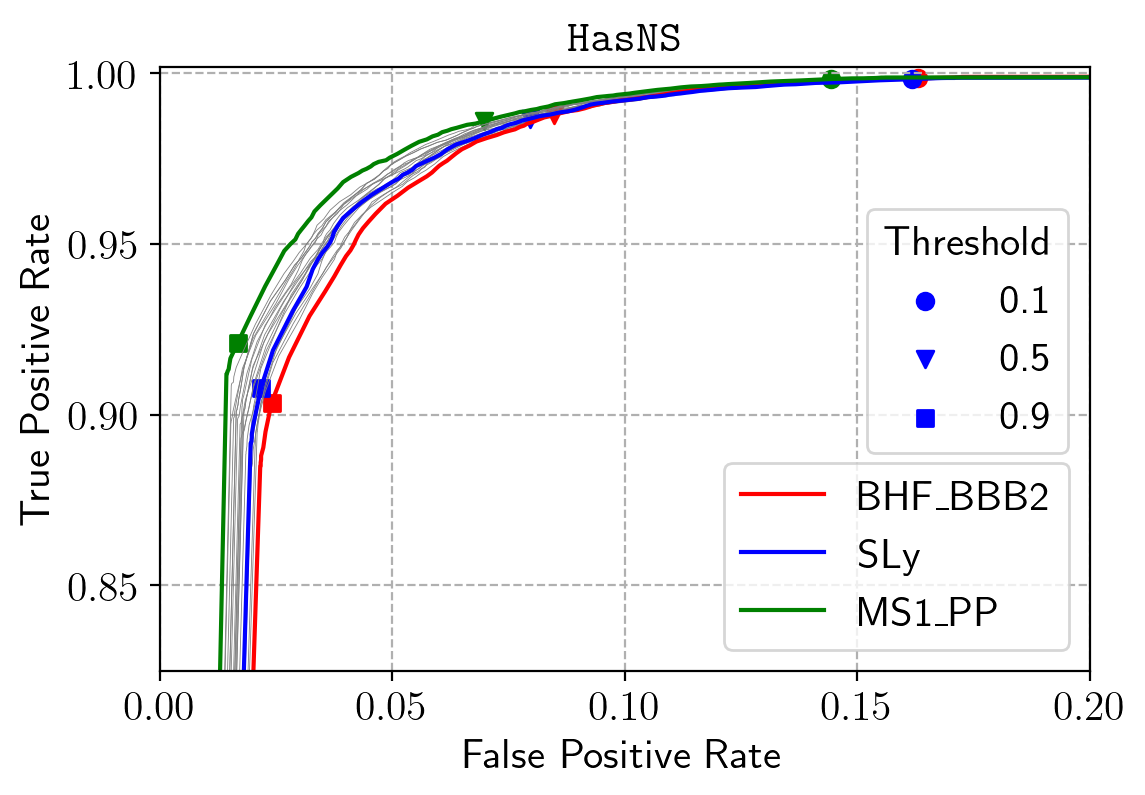}
\includegraphics[width=0.45\linewidth]{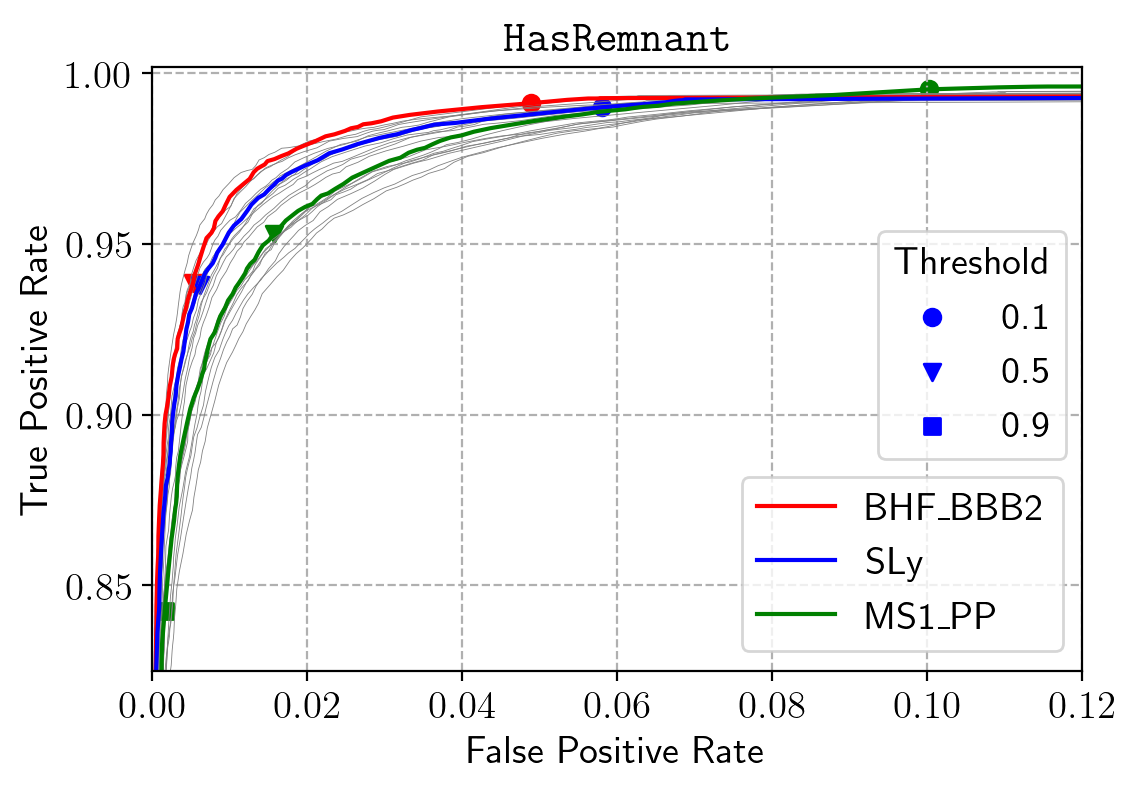}
\caption{\ac{ROC} curves obtained from the \ac{O2} testing data set $D_S$ for the \ac{KNN} classifier (left: \hasns, right: \hasrem). The curves for the 23 different \ac{EOS}s are displayed in
gray, with the curves for {\tt BHF\_BBB2}, {\tt MS1\_PP}, and {\tt SLy} highlighted in red, green, and blue, respectively. The circle, triangle, and square markers denote score thresholds of
$0.1$, $0.5$, and $0.9$, respectively.}
\label{fig:rocO2_KNN}
\end{figure*}

\begin{figure*}
\includegraphics[width=0.45\linewidth]{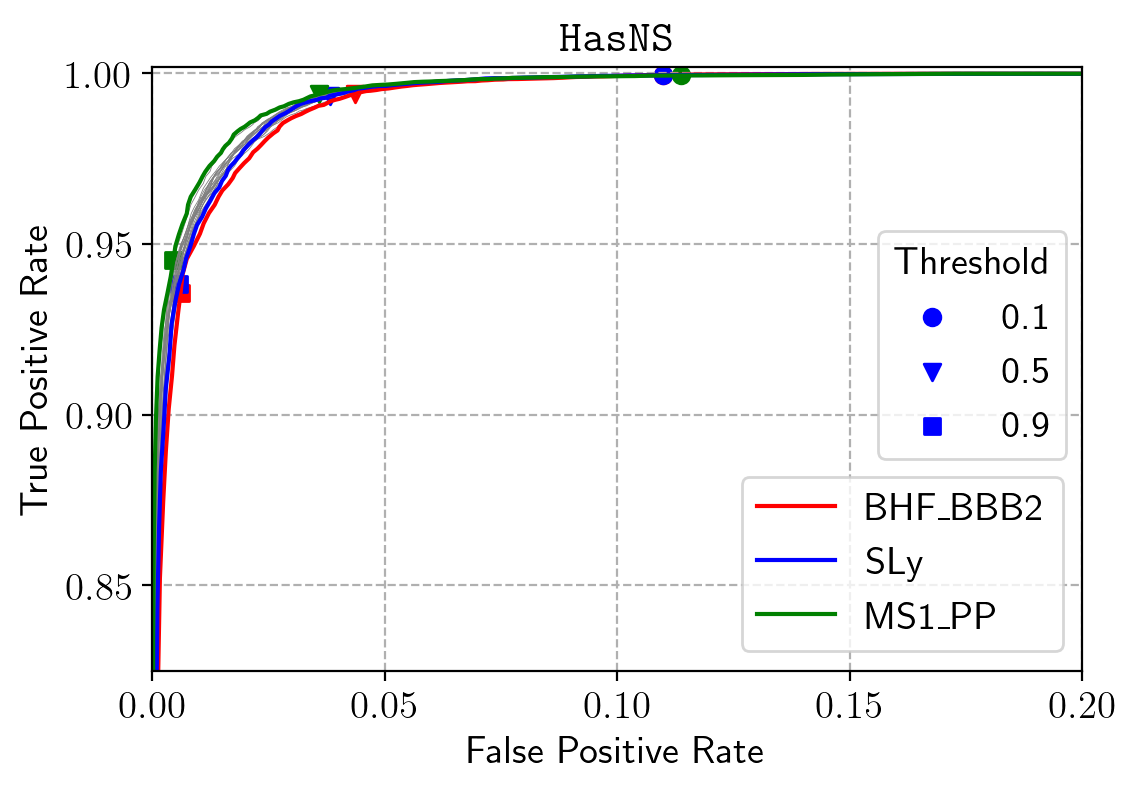}
\includegraphics[width=0.45\linewidth]{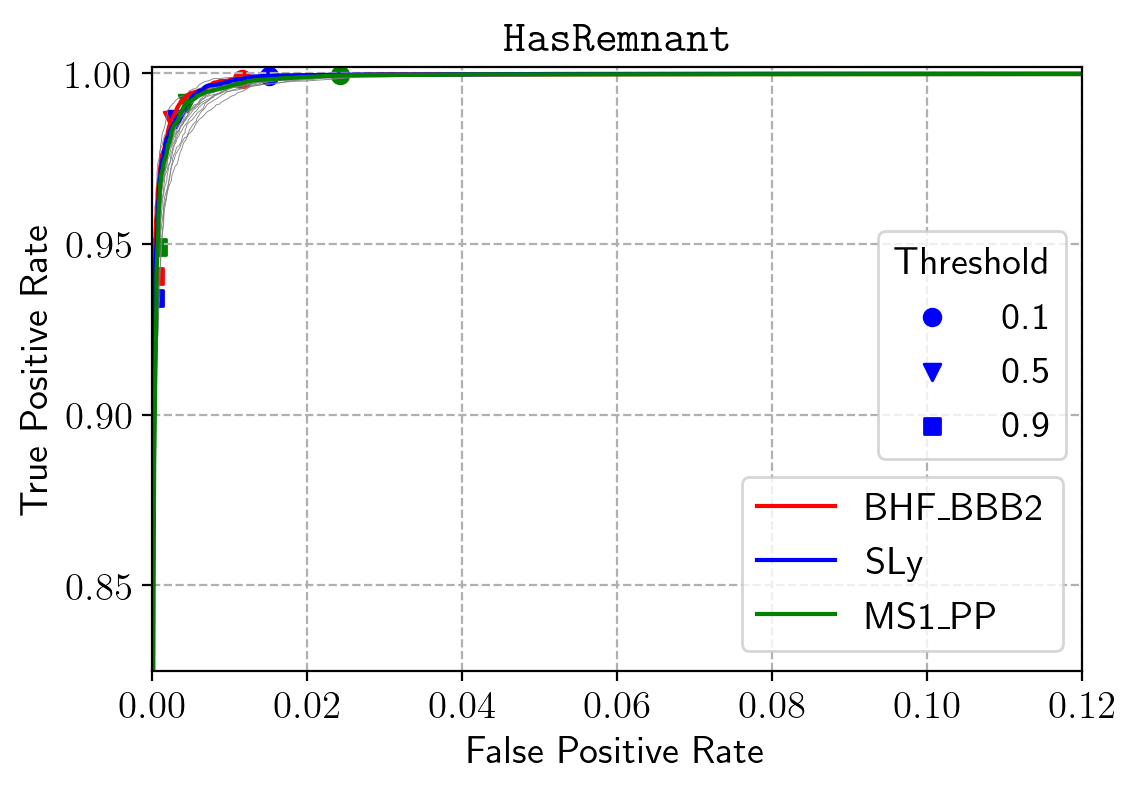}
\caption{\ac{ROC} curves obtained from the \ac{O2} testing data set $D_S$ for the \ac{RF} classifier (left: \hasns, right: \hasrem). The curves for the 23 different \ac{EOS}s are displayed in
gray, with the curves for {\tt BHF\_BBB2}, {\tt MS1\_PP}, and {\tt SLy} highlighted in red, green, and blue, respectively. The circle, triangle, and square markers denote score thresholds of
$0.1$, $0.5$, and $0.9$, respectively.}
\label{fig:rocO2_RF}
\end{figure*}

The \hasns\ and \hasrem\ \ac{ROC} curves for the  \ac{KNN} algorithm are displayed in the left and right panels of Fig.~\ref{fig:rocO2_KNN}, respectively. Figure \ref{fig:rocO2_RF}
displays the analogous curves for the \ac{RF} classifier. The \ac{ROC} curves for the 23 \ac{EOS} are plotted in gray, with three of them highlighted in color: {\tt BHF\_BBB2}, the
\ac{EOS} with the lowest maximum mass for the NS; {\tt MS1\_PP}, the \ac{EOS} with the largest maximum mass for the \ac{NS}; and {\tt SLy}, which allows for a maximum mass of $2.05
M_\odot$ and is the standard \ac{EOS} used in \ac{LVK} low-latency investigations~\cite{Ghosh:2021eqv}. The markers denote different thresholds for the algorithm scores. 

The two classifiers perform consistently across all \ac{EOS}s. The \ac{TPR} for a score threshold of $0.5$ is around $0.99$ for both \hasns\ and \hasrem. A comparison of the \hasns\ and
\hasrem\ \ac{ROC} curves for each algorithm shows that the \ac{FPR} for \hasns\ is generally higher than the \ac{FPR} for \hasrem\ at a given threshold. Thus, the algorithms typically do
a better job in classifying \hasrem\ than \hasns. Separate comparisons of the \ac{KNN} and \ac{RF} \ac{ROC} curves for \hasns\ and \hasrem\ reveal that at a given threshold, \ac{RF}
produces slightly higher \ac{TPR} and lower \ac{FPR} than \ac{KNN}.

\subsection{Bayesian Probability Computation}

After the algorithms have been trained and tested, we compute the Bayesian probabilities defined in Sec.~\ref{bayesian_probs} in terms of their outcomes. Equations~\eqref{bayes-hasns} and
\eqref{bayes-hasrem} must be assessed on a data set independent of the training data set, as stated in Sec.~\ref{bayesian_probs}. To do so, we employ $D_S$.

\begin{figure*}
\includegraphics[width=0.45\linewidth]{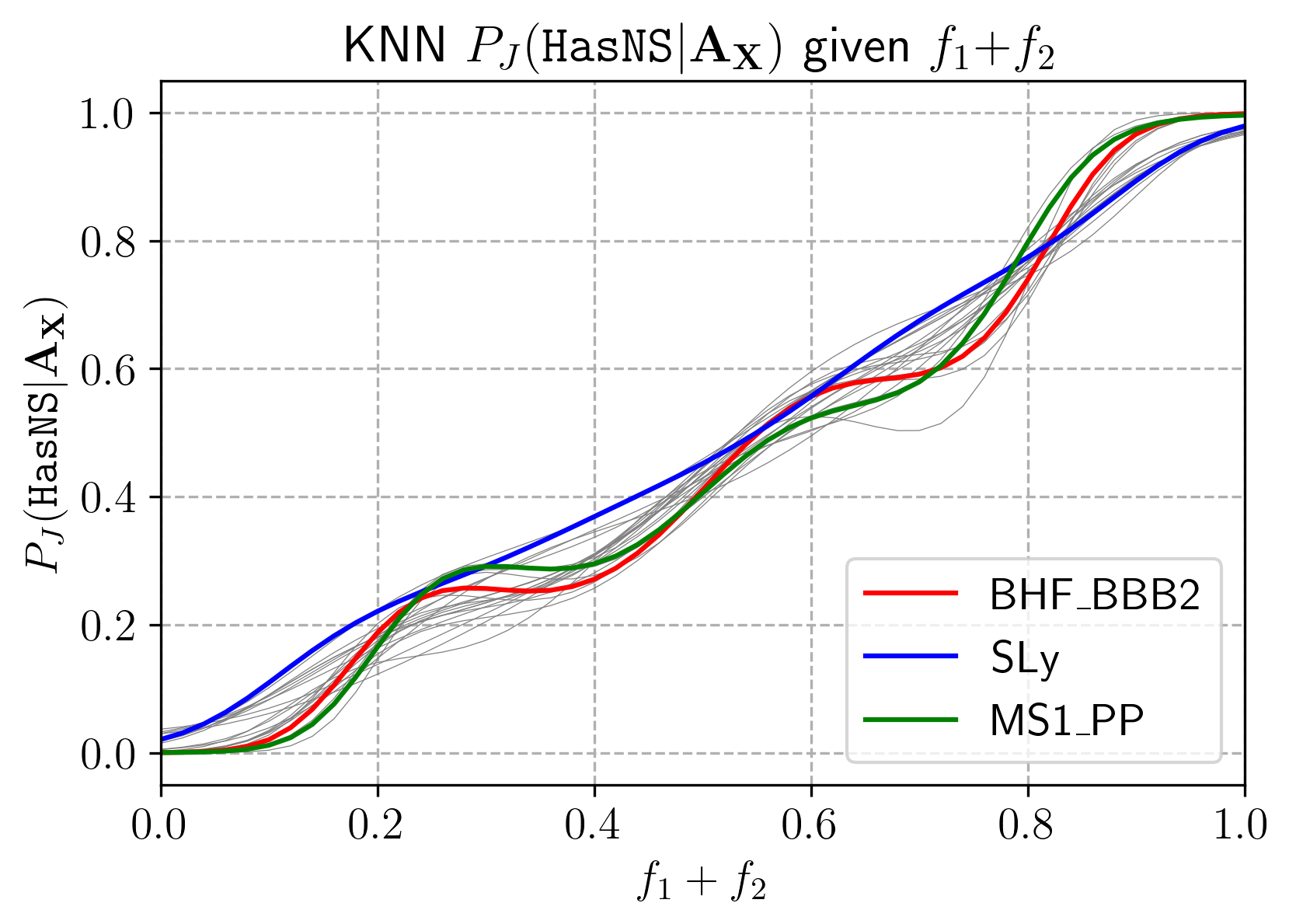}
\includegraphics[width=0.45\linewidth]{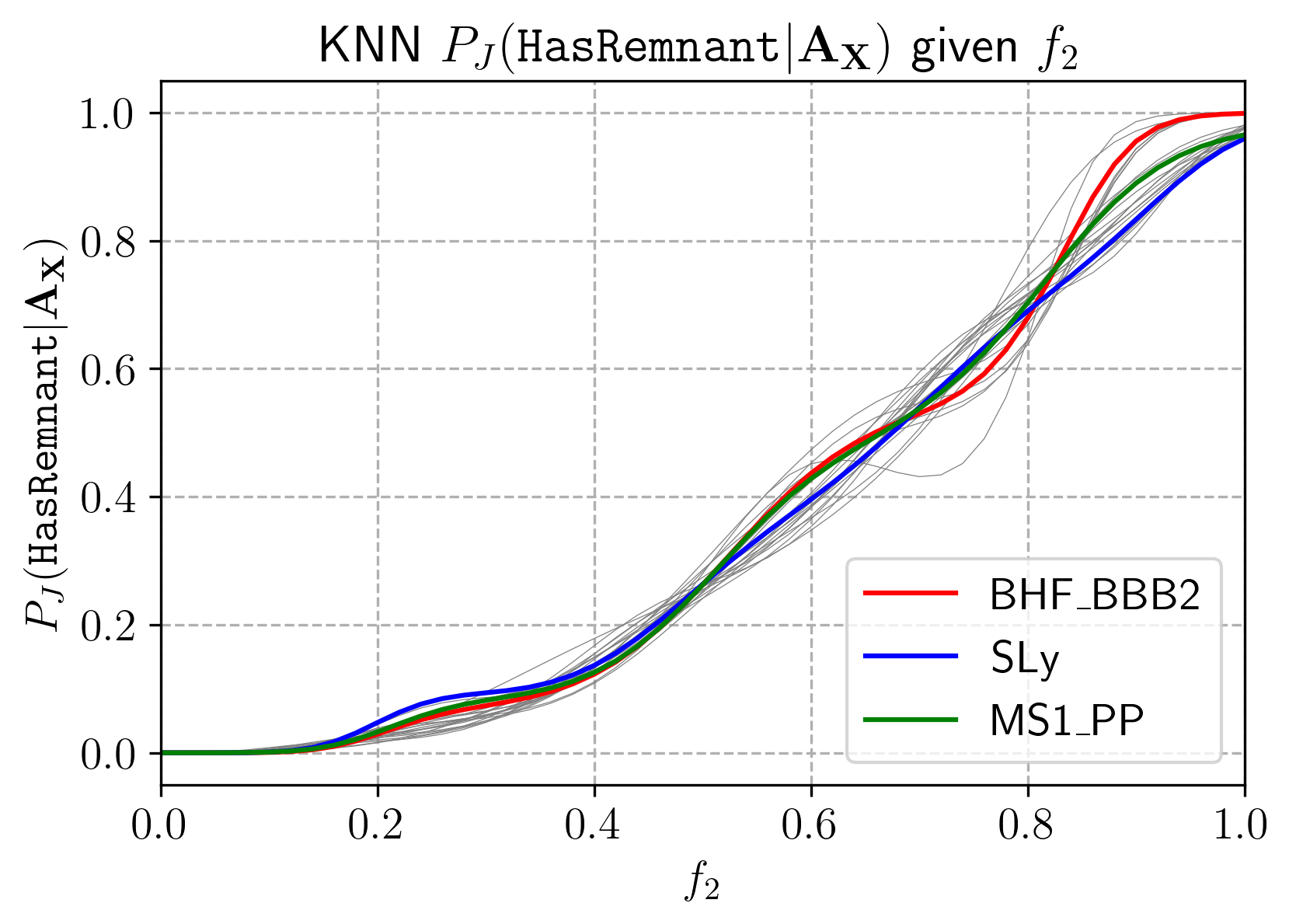}
\caption{Left panel: \hasns\ Bayesian probability curves for the 23 \ac{EOS} as a function of the fraction of \ac{KNN} neighbors $f_1+f_2$. Right panel: \hasrem\ Bayesian probability curves as a function of the fraction of \ac{KNN} neighbors $f_2$. Curves for the {\tt BHF\_BBB2}, {\tt MS1\_PP}, and {\tt SLy} \ac{EOS}s are highlighted in red, green, and blue, respectively. The probabilities show an increasing trend as the fraction of neighbors increases. Non-monotonic fluctuations are due to the data set's finite size.}
\label{fig:bayesian_prob_fits_KNN}
\end{figure*}

\begin{figure*}
\includegraphics[width=0.45\linewidth]{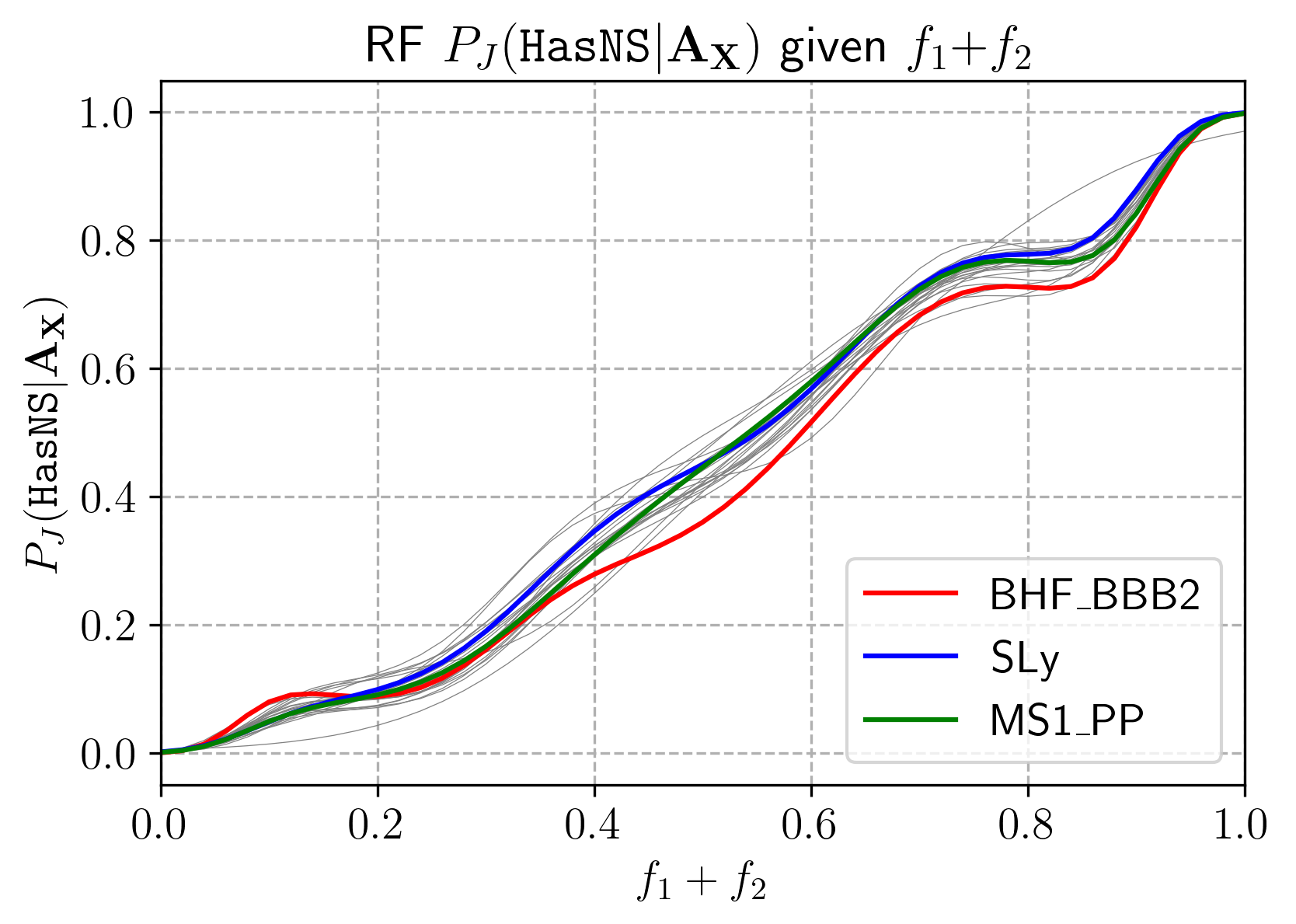}
\includegraphics[width=0.45\linewidth]{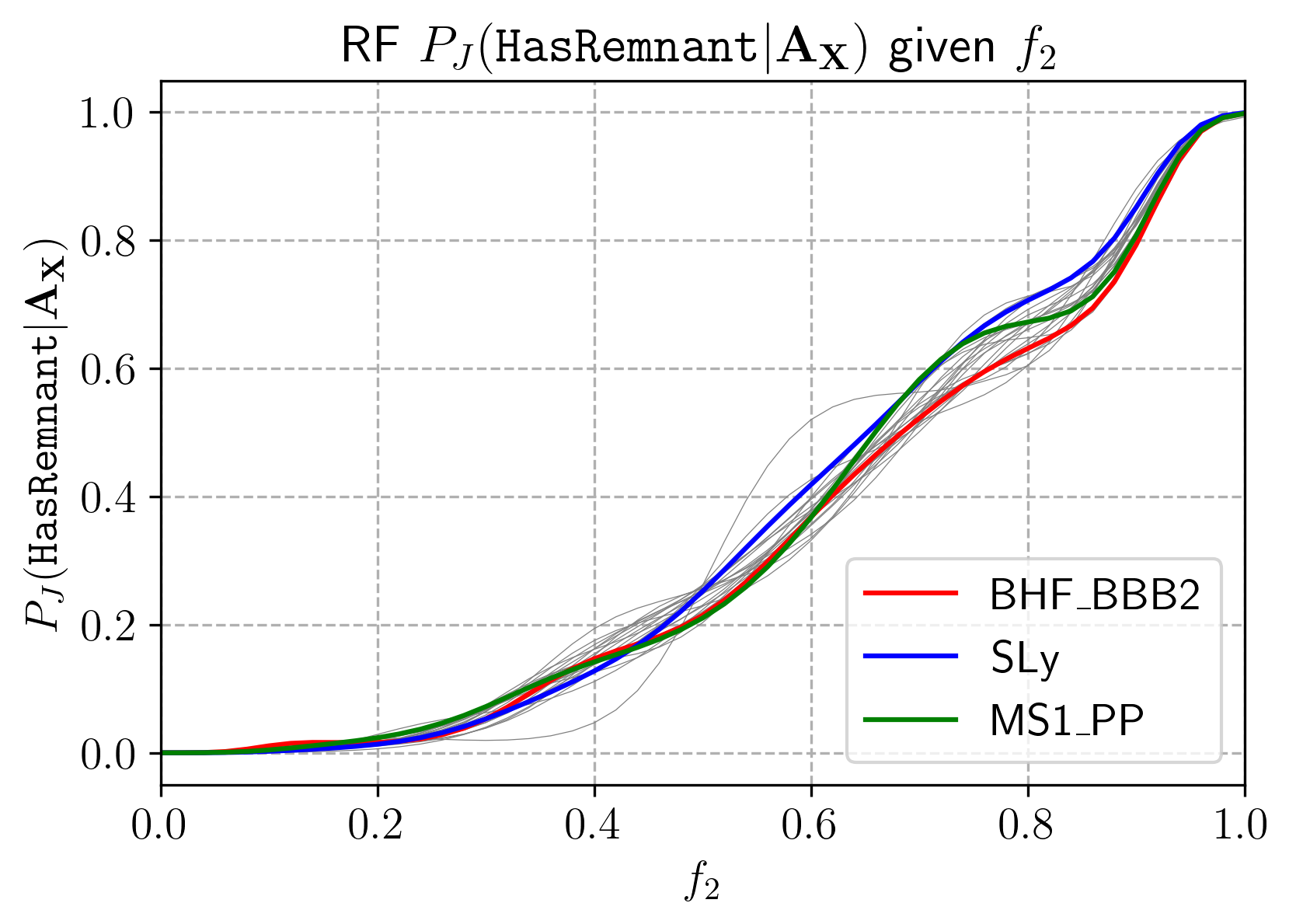}
\caption{Left panel: \hasns\ Bayesian probability curves for the 23 \ac{EOS} as a function of the fraction of \ac{RF} trees $f_1+f_2$. Right panel: \hasrem\ Bayesian probability curves for \hasrem\ as a function of the fraction of \ac{RF} Trees $f_2$. Curves for the {\tt BHF\_BBB2}, {\tt MS1\_PP}, and {\tt SLy} \ac{EOS}s are highlighted in red, green, and blue, respectively. The probabilities show an increasing trend as the fraction of trees increases.  Non-monotonic fluctuations are due to the data set's finite size.}
\label{fig:bayesian_prob_fits_RF}
\end{figure*}

The \ac{KNN} and \ac{RF} probability estimators are shown for each of the 23 \ac{EOS} in Figs.~\ref{fig:bayesian_prob_fits_KNN} and~\ref{fig:bayesian_prob_fits_RF}, respectively. As
expected, the \hasns\ and \hasrem\ probabilities increase with the fractions of \ac{KNN} neighbors (\ac{RF} trees). Local fluctuations in the probabilities are due to noise arising from
the finiteness of the data set. 

The probabilities in Figs.~\ref{fig:bayesian_prob_fits_KNN} and~\ref{fig:bayesian_prob_fits_RF} can be tabulated and used to compute the marginalized probabilities $P_M(\hasns|{\bf
A}_{\bf E})$ and $P_M(\hasrem|{\bf A}_{\bf E})$ as in Eq.~\eqref{bayes-marginalized}. 

\subsection{Bayesian Probabilities for the O3 Sets}

\begin{figure*}
\includegraphics[width=0.45\linewidth]{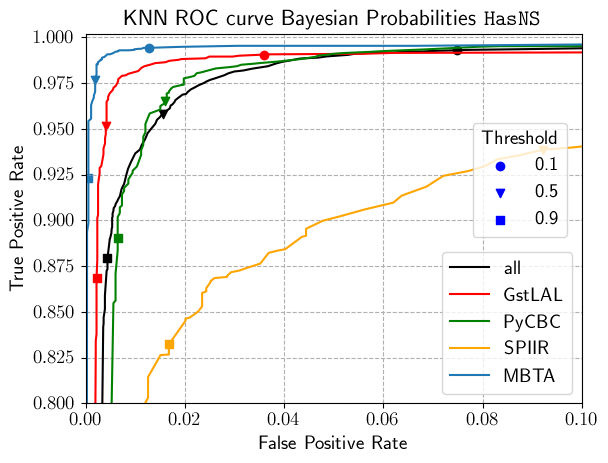}
\includegraphics[width=0.45\linewidth]{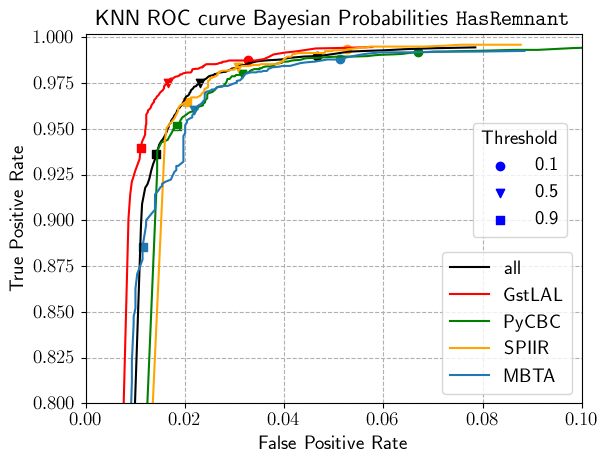}
\caption{\ac{ROC} curves obtained from the \ac{O3} \ac{MDC} data set for the \ac{KNN} classifier (left: \hasns, right: \hasrem). The different \ac{LVK} matched-filtering pipelines are indicated by different colors (GstLAL: red; PyCBC: green; gold: SPIIR; blue: MBTA). The results for all pipelines are shown in black. The circle, triangle, and square markers denote probability thresholds of $0.1$, $0.5$, and $0.9$, respectively.}
\label{fig:rocMDC_KNN}
\end{figure*}

\begin{figure*}
\includegraphics[width=0.45\linewidth]{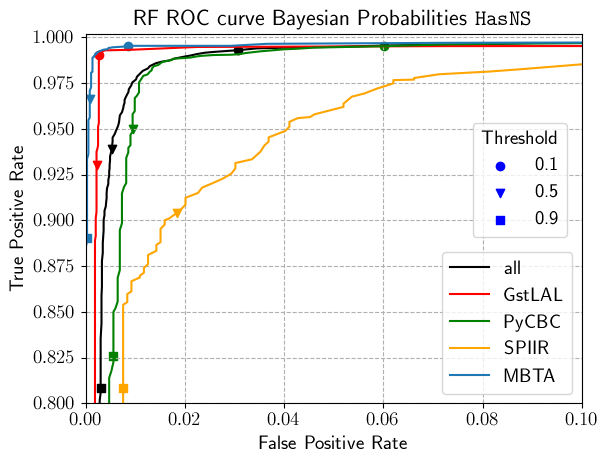}
\includegraphics[width=0.45\linewidth]{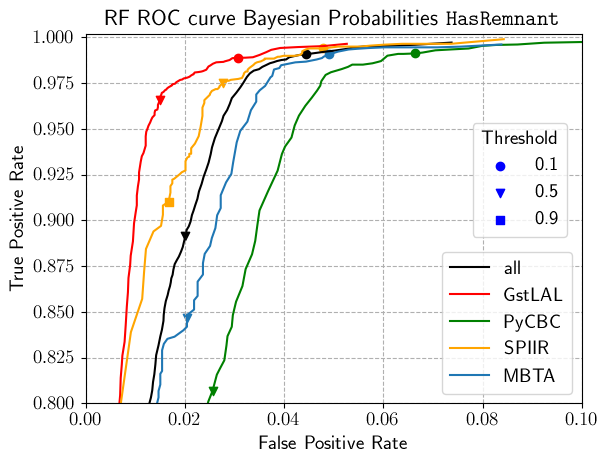}
\caption{\ac{ROC} curves obtained from the \ac{O3} \ac{MDC} data set for the \ac{RF} classifier (left: \hasns, right: \hasrem). The different \ac{LVK} matched-filtering pipelines are indicated by different colors (GstLAL: red; PyCBC: green; gold: SPIIR; blue: MBTA). The results for all pipelines are shown in black. The circle, triangle, and square markers denote probability thresholds of $0.1$, $0.5$, and $0.9$, respectively.}
\label{fig:rocMDC_RF}
\end{figure*}

To evaluate the method, we classify the events from the \ac{MDC} data set and compute the \ac{ROC}
curves based on the ground truth using Bayesian probability (rather than score) thresholds. The \ac{ROC} curves are shown in Figs.~\ref{fig:rocMDC_KNN} and \ref{fig:rocMDC_RF} for
\ac{KNN} and \ac{RF}, respectively. In contrast to the \ac{O2} data set, the \ac{MDC} set contains outputs from four matched-filtering pipelines
(GstLAL~\cite{Sachdev:2019vvd,PhysRevD.95.042001,Sachdev:2020lfd},  PyCBC~\cite{Nitz:2018rgo,DalCanton:2020vpm}, SPIIR~\cite{Chu:2020pjv}, and MBTA~\cite{Adams:2015ulm}). Therefore, we
present separate \ac{ROC} curves for these pipelines. 

In the case of \hasns, \ac{KNN} yields a \ac{TPR} between 0.95 and 0.98 and an \ac{FPR} smaller than 0.20 for a probability threshold of 0.5 across all pipelines, with the exception of
SPIIR. Even though the \ac{ML} algorithms are generally portable across pipelines and data sets, accurate results critically depend on the training set's faithful representation of the observations. The pipelines’ sub optimal performance, particularly SPIIR, can be explained by the fact that we only employed GstLAL to train the algorithms. SPIIR's interesting performance can also be seen in current LVK implementation~\cite{Chaudhary:2023vec}. Unfortunately, the inclusion of the other pipelines in the training phase requires the generation of additional injections by the LVK and is deferred to a future publication. Despite the restriction on GstLAL triggers, our approaches perform exceptionally well when applied to other pipelines. We believe that an increased injection set will further improve the performance of all pipelines.  

In regard to \hasrem, \ac{KNN} yields a \ac{TPR} around 0.975 and an \ac{FPR} slightly higher than 0.2 for the same threshold across all pipelines, with the exception of GstLAL. \ac{RF}'s results are similar to \ac{KNN}'s results. The \ac{RF} \ac{ROC} curves for \hasns\ typically have steeper slopes than for \ac{KNN}, resulting in a comparable \ac{TPR} but a lower \ac{FPR} at a given threshold. In the case of \hasrem, \ac{RF} performs similarly to \ac{KNN} for GstLAL but worse for the other pipelines.

A few interesting results are worth mentioning.  On the \ac{MDC} set, both algorithms perform better for \hasns\ than \hasrem, whereas on the \ac{O2} set, the reverse is true (see
Figs.~\ref{fig:rocO2_KNN} and~\ref{fig:rocO2_RF}). \ac{KNN} performs better than \ac{RF} on \hasrem, but does worse on \hasns. However, on events recovered by GstLAL, the pipeline on
which the algorithms have been trained, both algorithms exhibit comparable (high) performance. This seems to indicate that when used with other pipelines, \ac{RF} is less flexible than
\ac{KNN}. The conclusions provide more details about this and a possible explanation for this effect.

Finally, we apply the method to derive Bayesian probabilities for the events in the \ac{LVK} \ac{GWTC}~\cite{2021SoftX..1300658A,2023ApJS..267...29A}. In Table~\ref{tab:real_data_bayesian} we
report the results for some of the most significant \ac{GWTC} events, labeled with their event ID, and with their \hasns\ and \hasrem\ probabilities given in their \ac{GCN} Circulars\footnote{\url{https://gcn.nasa.gov}}. The $P_M(\hasns |\bf{A_E})$ and $P_M(\hasrem |\bf{A_E})$ probabilities for GW170817 and
GW190425, the two confirmed \ac{BNS} detections are $\sim 1$ as expected.  The probabilities for GW190426 and GW200115 (\ac{NSBH} mergers) are $P_M(\hasns |{\bf{A_E}}) \sim 1$ and
$P_M(\hasrem |{\bf{A_E}}) < 10^{-3}$. The two remaining significant events with non-zero probabilities are GW190814 and GW190924. These events were reported as high mass-ratio BBH
mergers. 

The fact that the system's component masses differ greatly from one another can be used to explain why $P_M(\hasns |\bf{A_E})$ for these events is not zero. In particular, the discrepancy
between \ac{RF} and \ac{KNN} for GW190814 can be understood from the different ways the two algorithms operate. \ac{RF} applies hard cuts on decision trees to evaluate its outcome.
\ac{KNN} looks at the fractions of neighbors surrounding the event. The detection pipeline returned a secondary mass compatible with an \ac{NS} for three of the 23 \ac{EOS}. However,
since the region of the parameter space close to the mass gap, i.e., the region between high \ac{NS} masses and low \ac{BH} masses, is not well covered in the \ac{O2} training data set,
\ac{KNN} overestimates the effect of the three \ac{EOS} predicting a secondary mass in the \ac{NS} region. 

\begin{table*}[]
\begin{tabular}{c|cc|cc|cc}
\hline
\multicolumn{1}{c|}{}  &  \multicolumn{2}{c|}{GCN Circular}                                                & \multicolumn{2}{c|}{$P_M(\hasns|{\bf A}_{\bf E})$}                                                & \multicolumn{2}{c}{$P_M(\hasrem|{\bf A}_{\bf E})$}                                                \\ \hline
\multicolumn{1}{c|}{event ID}  & \multicolumn{1}{c}{$P(\hasns)$} & \multicolumn{1}{c}{$P(\hasrem)$}  & \multicolumn{1}{c}{RF} & \multicolumn{1}{c}{KNN}  & \multicolumn{1}{c}{RF} & \multicolumn{1}{c}{KNN} \\ \hline
GW170817~\cite{2017GCN.21509....1L}      &  -- &          --                  & $>0.99$                 & 0.99                    & $>0.99$                   & 0.99                                \\
GW190425~\cite{2019GCN.24168....1L}     & $>0.99$      & $>0.99$                        & $>0.99$                  & 0.99                    & $>0.99$                   & 0.99                          \\
GW190426~\cite{2019GCN.24237....1L}                               &     $>0.99$      & $>0.99$ & $>0.99$                   & 0.99                    & $< 0.01$             & $< 0.01$                    \\
GW190814~\cite{2019GCN.25324....1L}                                  &   $< 0.01$ & $<0.01$ & 0.04                   & 0.57                   & $< 0.01$              & $< 0.01$                      \\
GW190924~\cite{2019GCN.25829....1L}                                                          &   $0.30$ & $<0.01$ & 0.01                  & 0.05                   & $< 0.01$              & $< 0.01$                       \\               
GW200115~\cite{2020GCN.26759....1L}                                                         &     $>0.99$      & $0.09$ & $>0.99$                   & 0.99                   & $< 0.01$              & $< 0.01$                           \\
\hline
\end{tabular}
\caption{Bayesian probabilities of a few significant \ac{GW} events from the \ac{GWTC} catalog. GW170817 and GW190425, GW190426 and GW200115, and GW190814 and GW190924 were determined in \ac{LVK}'s follow-up investigations to be \ac{BNS}, \ac{NSBH}, and \ac{BBH} mergers, respectively. Probabilities for GW170817 are not reported as the event was observed by \ac{EM} observatories and its true nature is confirmed. The probability values in the table are rounded to two decimal figures.}
\label{tab:real_data_bayesian}
\end{table*}

\begin{figure*}
\includegraphics[width=0.9\linewidth]{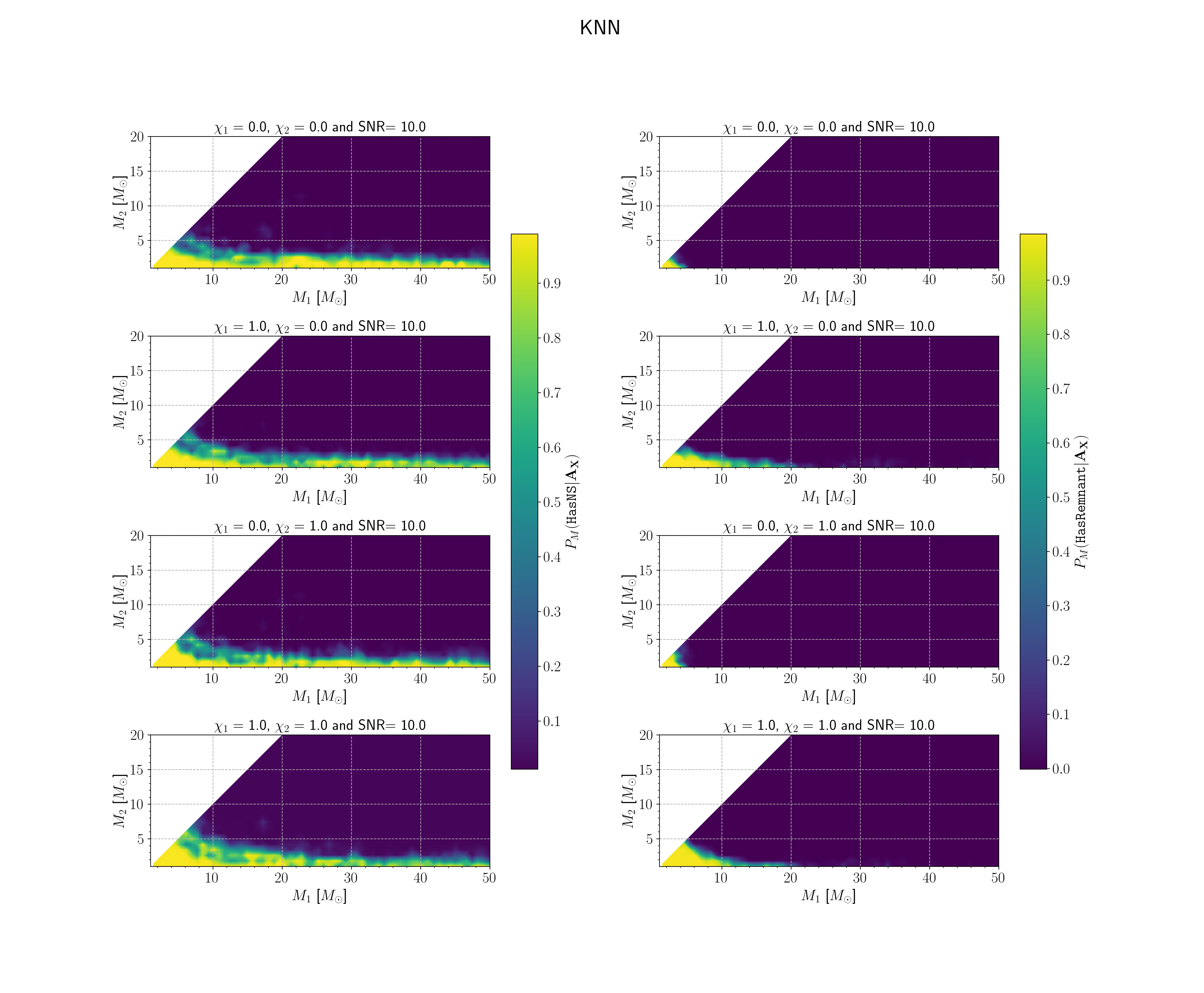}
    \caption{Parameter sweeps for $P_M(\hasns |\bf{A_E})$ (left panels) and $P_M(\hasrem |\bf{A_E})$ (right panels) for the \ac{KNN} algorithm. $M_1$ and $M_2$ are the primary and secondary component masses of the binary. $\chi_1$ and $\chi_2$ are their effective spins. The \ac{SNR} is fixed to 10.}
\label{fig:param_sweep_KNN}
\end{figure*}

\begin{figure*}
\includegraphics[width=0.9\linewidth]{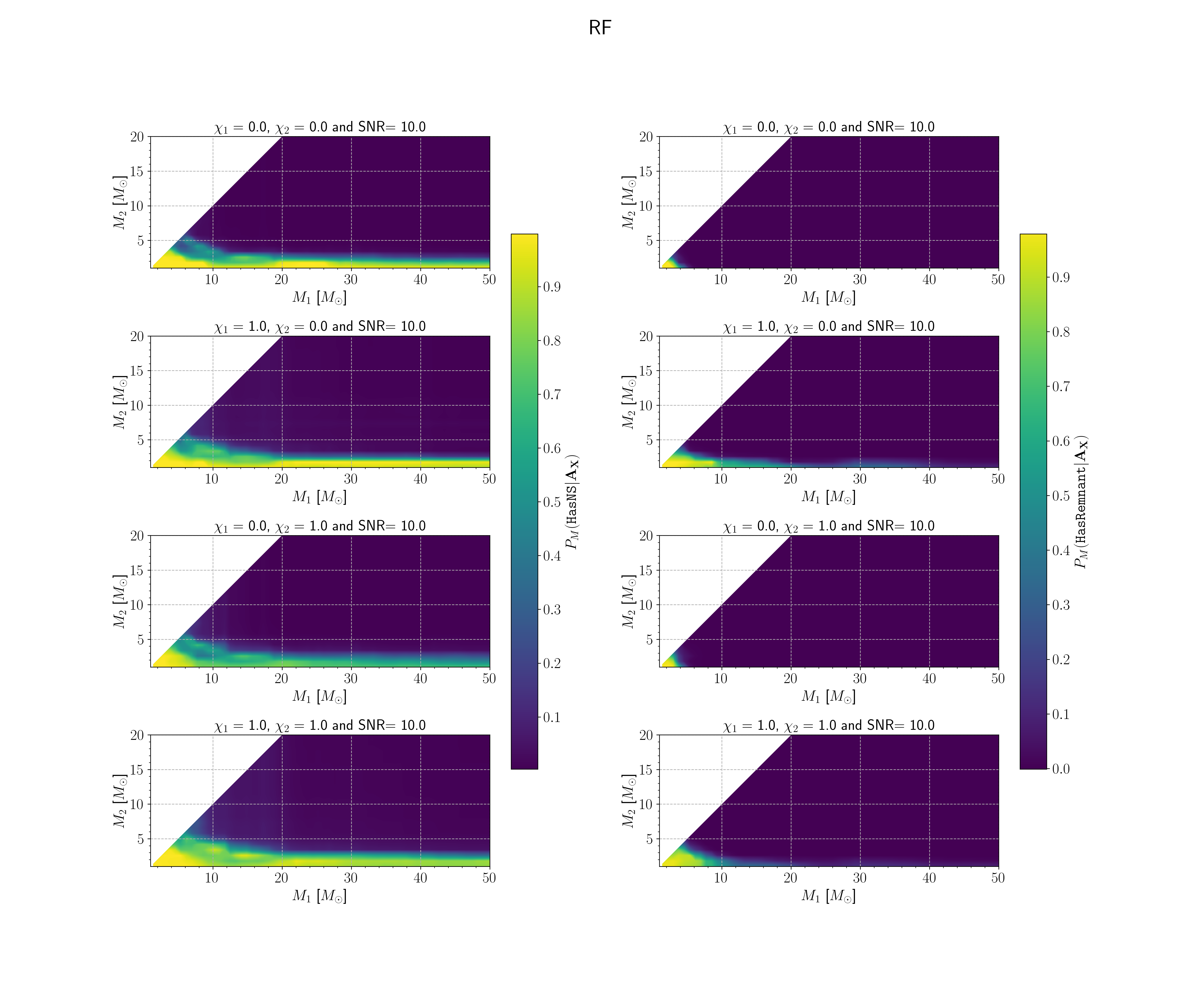}
    \caption{Parameter sweeps for $P_M(\hasns |\bf{A_E})$ (left panels) and $P_M(\hasrem |\bf{A_E})$ (right panels) for the \ac{RF} algorithm. $M_1$ and $M_2$ are the primary and secondary component masses of the binary. $\chi_1$ and $\chi_2$ are their effective spins. The \ac{SNR} is fixed to 10.}
\label{fig:param_sweep_RF}
\end{figure*}

Figures~\ref{fig:param_sweep_KNN} and~\ref{fig:param_sweep_RF} show parameter sweeps in the space of the binary component masses for the \ac{KNN} and \ac{RF} Bayesian probabilities,
respectively. Different rows correspond to different values of the component spins. Both algorithms perform in a similar way for $P_M(\hasns |\bf{A_E})$. However, the parameter sweeps for
$P_M(\hasns |\bf{A_E})$ for \ac{KNN} are noisier than \ac{RF} for large primary masses. As was noted above, the \ac{KNN} algorithm operates by looking at the closest neighbors. If
neighbors with different labels are present in the region of interest, the outcome is bound to be noisy. The \ac{RF} algorithm applies hard selection cuts to primary masses. This results
in a more uniform probability. Changes in spin seem to not significantly affect the outcome.  Similar behaviors for \ac{KNN} and \ac{RF} can also be observed in the case of $P_M(\hasrem
|\bf{A_E})$. As expected from the Foucart formula, $P_M(\hasrem |\bf{A_E})$ increases with the primary mass for large primary spins, and the region with $P_M(\hasrem |\bf{A_E}) \sim 1$ is
included in the region where $P_M(\hasns |\bf{A_E}) \sim 1$.

\section{Conclusions\label{conclusions}}

In this paper, we have presented a new scheme for real-time classification of \ac{GW} \ac{CBC} signals detected by the \ac{LVK} detectors. The method uses the output of the \ac{LVK}
low-latency pipelines to identify whether the GW source progenitor contains a \ac{NS} (\hasns) and a post-merger matter remnant is produced in the merger (\hasrem). Estimates of these
metrics are included in public alerts for candidate \ac{GW} events issued by the \ac{LVK}. Determining these metrics in low latency is crucial to enabling coincident \ac{MMA} observations
of GW and \ac{EM} signatures.

We have assessed the viability and measured the performance of two classifiers, \ac{KNN} and \ac{RF}, on two sets of real detector data augmented with synthetic GW injections of GW
signals that were generated for space-time volume sensitivity analyses of \ac{O2} \ac{LVK} \ac{GW} searches~\cite{Chatterjee:2019avs} and an \ac{MDC} real-time replay of \ac{O3}
data~\cite{Chaudhary:2023vec}.

One important novel ingredient of the proposed scheme is the computation of Bayesian probabilities for \hasns\ and \hasrem. Until now, the information that has been passed to astronomers
in public alerts has been in the form of binary classification scores for these metrics. Here, we provide a method to compute \hasns\ and \hasrem\ as actual probabilities that the \ac{GW}
source includes a neutron star and post-merger matter remnant. Therefore, our scheme provides more direct and easily interpretable information to aid the community of astronomers in
deciding whether to follow up on \ac{GW} candidate events with \ac{EM} observatories.

To construct the Bayesian probabilities for \hasns\ and \hasrem, we train and test the classifiers on the \ac{O2} data set following the customary 70\% -- 30\% split between training and
testing data. After evaluating the performance of the classifiers with standard \ac{ROC} curves, we use the testing set to generate numerical Bayesian probability expressions for the
models. This minimizes potential bias that may result from the use of data sets with different properties while ensuring that the Bayesian fits are built with data that is independent of the data used for training the classifiers. The effectiveness of the Bayesian fits is then evaluated on fully independent data sets using the \ac{O3} set and real detections. 

As shown in Appendix~\ref{app:comparison}, our \ac{RF} implementation outperforms the specific \ac{KNN} algorithm implementation currently utilized in the \ac{LVK} low-latency infrastructure, while our \ac{KNN} method performs similarly. When tested on the \ac{O3} set, both algorithms improve on \hasns\ while underperforming on
\hasrem.  In this case, \ac{KNN} outperforms \ac{RF}, which exhibits more variation across different pipelines.  If only the injections recovered by GstLAL are considered in \ac{O3}, the
\ac{O3} results of both \ac{RF} and \ac{KNN} are consistent with \ac{O2}. \ac{RF}'s performance on \ac{O3} events recovered by other pipelines, on the other hand, is noticeably lower.
This appears to imply that \ac{RF} is less portable than \ac{KNN} across different data sets and pipelines. The different ways \ac{RF} and \ac{KNN} operate, as well as the different
characteristics of the sets, may explain their behavior.

The RF classifier is a decision tree-based classifier that sets decision rules by implementing specific cuts (conditions) on input features. The \ac{KNN} algorithm implements decision
rules by computing the nearest neighbors of input features in the parameter space for the data point of interest. \ac{RF} is designed to construct hard boundaries based on input
parameters, whereas the \ac{KNN} algorithm is designed to produce an outcome based on differences between features. As a result, the \ac{RF} algorithm's nature may make it more suitable
for classifying events with \hasns, which is based on a well-defined, hard boundary between positive and negative outcomes, such as the secondary mass value. To distinguish between
systems with zero and nonzero post-remnant matter in \hasrem, the algorithms must learn Foucart's fit from the recovered parameters. Foucart's formula is dependent on the \ac{EOS} under
consideration, as well as the pipeline that recovers the injection. Because \ac{RF} and \ac{KNN} are trained on injections that are only recovered by GstLAL, \ac{RF} is more affected than
\ac{KNN}. However, a comprehensive evaluation, involving datasets from new observing runs and diverse pipelines, would be needed to definitely compare the performance of the two algorithms.

This work provides an improved scheme to implement Bayesian probabilities for \hasns\ and \hasrem\ classification of candidate events that would be straightforward to deploy in the
existing LVK infrastructure. Our method can also be easily extended to other properties of \ac{GW} signals that are being or may be released in low latency, such as \hasgap\ among other
data products. Other future extensions of this work include improving algorithm training and Bayesian fit estimation with updated data sets of simulated injections in \ac{LVK} \ac{O4}
data generated with different pipelines and with better coverage of the mass gap region than the \ac{O2} data set. It would also be worthwhile to investigate the use of additional ML
classifiers that could further improve the process's accuracy, reduce the need for computational resources, and decrease latency. Finally, a similar infrastructure could be designed and
deployed to aid in the rapid parameter estimation of pipeline outputs, but with a focus on feature regression rather than classification. This latter line of investigation will be
presented in a future publication.

\section*{Acknowledgments}

We are very grateful to the authors of Ref.~\cite{Chatterjee:2019avs} for fruitful discussions and for sharing their work that helped us to directly compare our methods to those used in
\ac{LVK}'s \ac{O3} observing run. We would also like to thank the many other colleagues of the LIGO Scientific Collaboration and the Virgo Collaboration who have provided invaluable help
over the years and, in partcular, Tito Dal Canton, Thomas Dent, and Shaon Ghosh for their useful comments during the internal \ac{LVK} review period.

This work is based upon work supported by the LIGO Laboratory which is a major facility fully funded by the National Science Foundation. We are grateful for computational resources
provided by the LIGO Laboratory and supported by the U.S.\ National Science Foundation Awards PHY-0757058 and PHY-0823459. 

M.B.\ is partiallly supported by the Spanish Agencia Estatal de Investigaci\'on grant PID2020-118236GB-I00. M.M.T.\ is partially supported by the Ministerio de Ciencia, Innovación y Universidades de Espa\~na through the ``Ayuda para la Formaci\'on de Profesorado Universitario'' (FPU) grant FPU19/01750, by the Spanish Agencia Estatal de Investigaci\'on grants
PGC2018-095984-B-I00 and PID2021-125485NB-C21 funded by MCIN/AEI/10.13039/501100011033 and the European Regional Development Fund (ERDF) ``A way of making Europe,'' and by the Generalitat
Valenciana (Prometeo grant CIPROM/2022/49). S.S.C.\ and Y.Z.\ are partially supported by the U.S.\ National Science Foundation under awards PHY-2011334 and PHY-2308693. M.C.\ is partially
supported by the U.S.\ National Science Foundation under awards PHY-2011334, PHY-2219212 and PHY-2308693. L.M.Z.\ is partially supported by the MSSGC Graduate Research Fellowship, awarded
through the NASA Cooperative Agreement 80NSSC20M0101. D.T.~is partially supported by the Spanish Agencia Estatal de Investigaci\'on grant PID2021-125485NB-C21 funded by
MCIN/AEI/10.13039/501100011033 and ERDF ``A way of making Europe,'' and by the Generalitat Valenciana (Prometeo grant CIPROM/2022/49).

M.W.C.\ acknowledges support from the National Science Foundation with grant numbers PHY-2308862 and OAC-2117997. A.T.\ acknowledges support from the National Science Foundation with
grant numbers PHY-2010970 and OAC-2117997.

This research was initiated by M.B, M.M.T., S.A., M.C., L.M.Z., D.T., and Y.Z.\ during their stay at the Institute of Pure and Applied Mathematics (IPAM),  University of California
Los-Angeles (UCLA). IPAM is partially supported by the National Science Foundation through award DMS-1925919. These authors would like to thank IPAM and UCLA for their warm hospitality.

Software for this analysis is written in \texttt{Python} 3.x~\cite{python3} and uses standard Open Source libraries and community-contributed modules from the Python Package Index (PyPI)
repository~\cite{pypi} including \texttt{numpy}~\cite{harris2020array}, \texttt{scipy}~\cite{2020SciPy-NMeth}, \texttt{pandas}~\cite{reback2020pandas,mckinney-proc-scipy-2010},
\texttt{matplotlib}~\cite{Hunter:2007ouj}, and \texttt{sklearn}~\cite{scikit-learn}.

This manuscript has been assigned LIGO Document Control Center number P2300338.

\textbf{Author contribution statement.} M.B.\ and M.M.T.\ designed the \ac{RF} and \ac{KNN} models and their computational framework, respectively, and ran the corresponding analyses.
S.S.C.\ curated the data sets and contributed to the analysis, particularly the comparison with the current \ac{LVK} \ac{KNN} implementation. M.C.\ directed the project. S.A., L.M.Z.,
D.T., and Y.Z.\ contributed to all aspects of this work, including but not limited to the design and implementation of the research, data analysis, and manuscript writing.  Authorship for
the \ac{LVK} members of the \ac{MDC} builder's list is requested in publications resulting from the \ac{MDC} data set.  M.W.C.\ and A.T.\ are opt-in members of this list. M.W.C.\ also
provided some comments during the internal \ac{LVK} review period. The authors gratefully thank the other opt-out members of the \ac{MDC} builder's list for creating the set, and/or
providing summary statistics based on its results, and making these products available to the \ac{LVK} collaborations: Sarah Antier, Patrick J.~Brockill, Deep Chatterjee, Reed Essick,
Shaon Ghosh, Patrick Godwin, Erik Katsavounidis, and Gaurav Waratkar.

\appendix

\section{Cross validation of the ML algorithms}  \label{app:crossval}

To obtain the optimal hyperparameters of our \ac{ML} algorithms, we perform cross-validation across different values of the algorithm parameters. We apply the algorithms to the \ac{O2} data set for different choices of parameters, and we select the combination that gives the highest accuracy.

\begin{figure*}
\includegraphics[width=\linewidth]{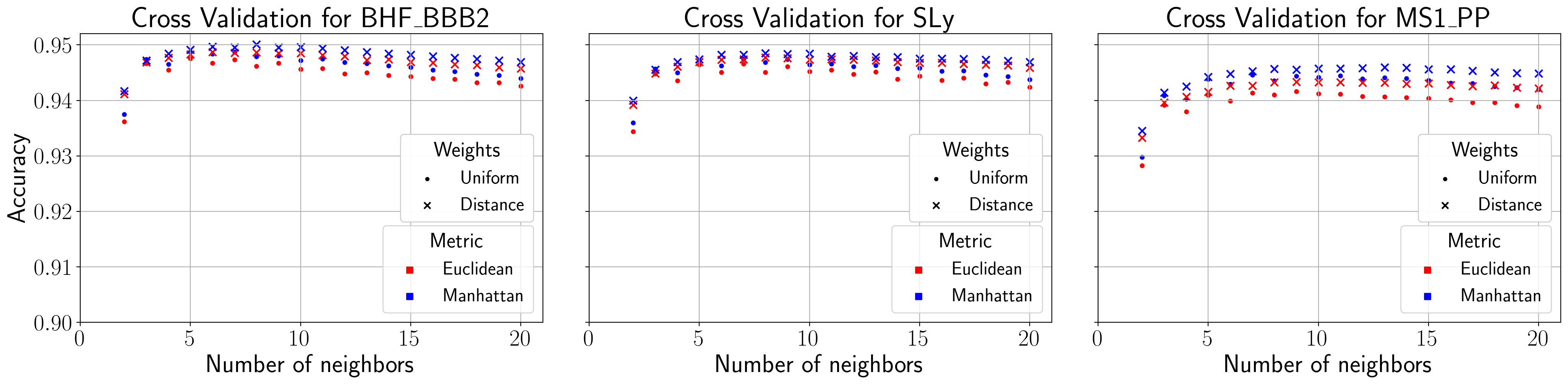}
\caption{Accuracy of the \ac{KNN} algorithm as a function of the number of neighbors. The color red (blue) denotes the Euclidean (Manhattan) metric, and the dots (crosses) denote uniform (distance) weights. The BallTree algorithm is used to find the nearest neighbors. The accuracy does not significantly change when more than 5 neighbors are considered.}
\label{fig:crossvalKNN}
\end{figure*}

\subsection{K-Nearest Neighbors}

We utilize the $k$-fold cross validation implemented in scikit-learn~\cite{Pedregosa:2011ork} with $k = 10$ folds, which is a common choice in applied machine learning. As consistency checks, we use different numbers of folds ($k = 4$ and $k=5$) and achieve consistent results in all cases, with an accuracy ranging between 0.9449 and 0.9517. The dataset used for cross-validation is $D=D_R\oplus D_S$.

The \ac{KNN} algorithm's hyperparameters include the number of neighbors ($K$), distance metric, nearest neighbor method, and prediction weight function. Table ~\ref{tab:KNN_crossval_tab} displays many combinations of these parameters and the resulting accuracy following cross-validation. It is worth noting that the choice of hyperparameters has little to no impact on accuracy.

\begin{table}[h]
\begin{tabular}{c|c|c|c|c}
\hline
Neighbors   & Metric & Weights & Algorithm & \multicolumn{1}{|c}{Accuracy} \\ \hline
2                                   & Manhattan                   & Distance                    & BallTree                   &  0.9399                                  \\
2                                   & Euclidean                   & Distance                    & KDTree                   & 0.9392                           \\
2                                   & Manhattan                   & Uniform                    & Auto             &                    0.9359  \\
8                                   & Manhattan                  & Distance                   & BallTree &                    0.9485 \\
8                                   & Euclidean                 & Distance                   & KDTree             &                       0.9477 \\               
8                                   & Manhattan                   & Uniform                   & Auto              &                          0.9468 \\
20                                   & Manhattan                   & Distance                   & BallTree &                  0.9469 \\
20                                   & Euclidean                   & Distance                   & KDTree              &                       0.9459 \\               
20                                   & Manhattan                   & Uniform                   & Auto              &                          0.9437 \\
\hline
\end{tabular}
\caption{Values of the accuracy for different combinations of the hyperparameters. The accuracy in all cases is larger than $0.935$.}
\label{tab:KNN_crossval_tab}
\end{table}

Figure~\ref{fig:crossvalKNN} illustrates the accuracy dependence on different hyperparameter options while keeping the algorithm used to compute nearest neighbors, the BallTree, constant.

The ideal hyperparameters are the same for all state equations, with the exception of the number of neighbors, which varies between 6 and 12.  We use the Bayes factors described in~\cite{Ghosh:2021eqv} to marginalize over all state equations. The marginalized optimal number of neighbors is 8. However, as illustrated in Fig.~\ref{fig:crossvalKNN}, there is no significant change in accuracy when a higher number of neighbors is used. Table~\ref{tab:KNN_opt_params} shows the fixed hyperparameters used in our KNN implementation.

\begin{table}[h]
\begin{tabular}{c|c|c|c}
\hline
Nº of neighbors   & Metric & Weights & Algorithm  \\ \hline
8  & Manhattan  & Distance & BallTree \\ 
\hline
\end{tabular}
\caption{Optimal hyperparameters for the \ac{KNN} algorithm. These are the values we use to train and test the algorithm, as well as to calculate Bayesian probabilities.} \label{tab:KNN_opt_params}
\end{table}

\subsection{Random Forest}

\begin{figure*}
\includegraphics[width=\linewidth]{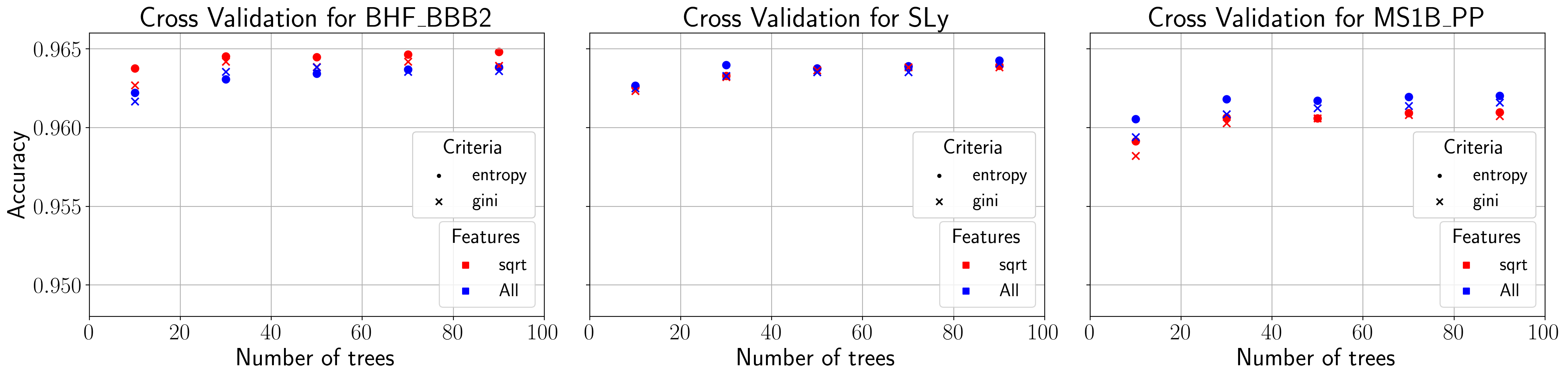}
\caption{Accuracy of the \ac{RF} algorithm as a function of the number of trees.  The color red (blue) denotes ``sqrt'' (``all'') features, while the dots (crosses) denote ``entropy'' (``gini'') criteria. The depth of the trees is set at 15. All cases have an accuracy range of 0.960 to 0.965.}
\label{fig:crossvalRF}
\end{figure*}

To apply cross-validation to the \ac{RF} algorithm, we use the training ($D_R$) and testing ($D_S$) data sets presented in Section~\ref{probability}. We choose possible values for the various hyperparameters and compare the accuracy of the generated forests. Unlike \ac{KNN}, \ac{RF} does not apply the $k$-fold cross-validation procedure. This is because the implementation of \ac{RF} we employ admits the bootstrap technique in the training. Each tree in the forest accesses a random subset of the training data, thus achieving the same effect.

We investigate the effect on accuracy by varying the number of trees in the forest, the method used to compute the information gain after each node split (gini or entropy), the maximum number of features considered in each node (the square root of the total number of features, or all of them), and the maximum depth that the trees can reach. For tree depth, we explore two options: None, which consists of allowing the trees to grow until data points are isolated at the leaves, and 15. We chose this figure since it is half the average depth obtained with the None option.

We use the Bayes factors to marginalize on the configurations that lead to the highest score for each \ac{EOS}. We conclude that for all \ac{EOS}, we should employ 81.017 trees, the ``entropy'' criterion, the square root of the number of features at split, and a depth equal to 15.

\begin{table}[h]
\begin{tabular}{|l|c|}
\hline
Trees& 81.02 \\
Criteria (0-gini, 1-entropy)& 0.9577 \\
Features (0-sqrt, 1-All)& 0.3902  \\
Depth (0-15 depth, 1-None)& 0.0000 \\ \hline
\end{tabular}
\caption{Optimal hyperparameters for the \ac{RF} algorithm. \label{tab:RF_cross_params}}
\end{table}

Table \ref{tab:RF_cross_params} displays the results of hyperparameter marginalization over the \ac{EOS}. For any \ac{EOS}, regardless of other hyperparameters, a depth of 15 yields higher accuracy. The "entropy" criterion is adopted for practically every \ac{EOS}, whereas the number of features to consider is more tightly restricted.

Figure~\ref{fig:crossvalRF} displays the accuracies obtained during cross-validation, with the depth fixed to 15. It is worth noting that using more trees results in higher scores. Nonetheless, the goal of making this method run in low latency favors smaller files (with less tree information) that can be loaded rapidly. Therefore, given the difference in accuracy and size of the trained models, we limit ourselves to 50 trees.

\section{Comparison between our algorithms and the current LVK implementation}  \label{app:comparison}

\begin{figure*}
\includegraphics[width=0.47\linewidth]{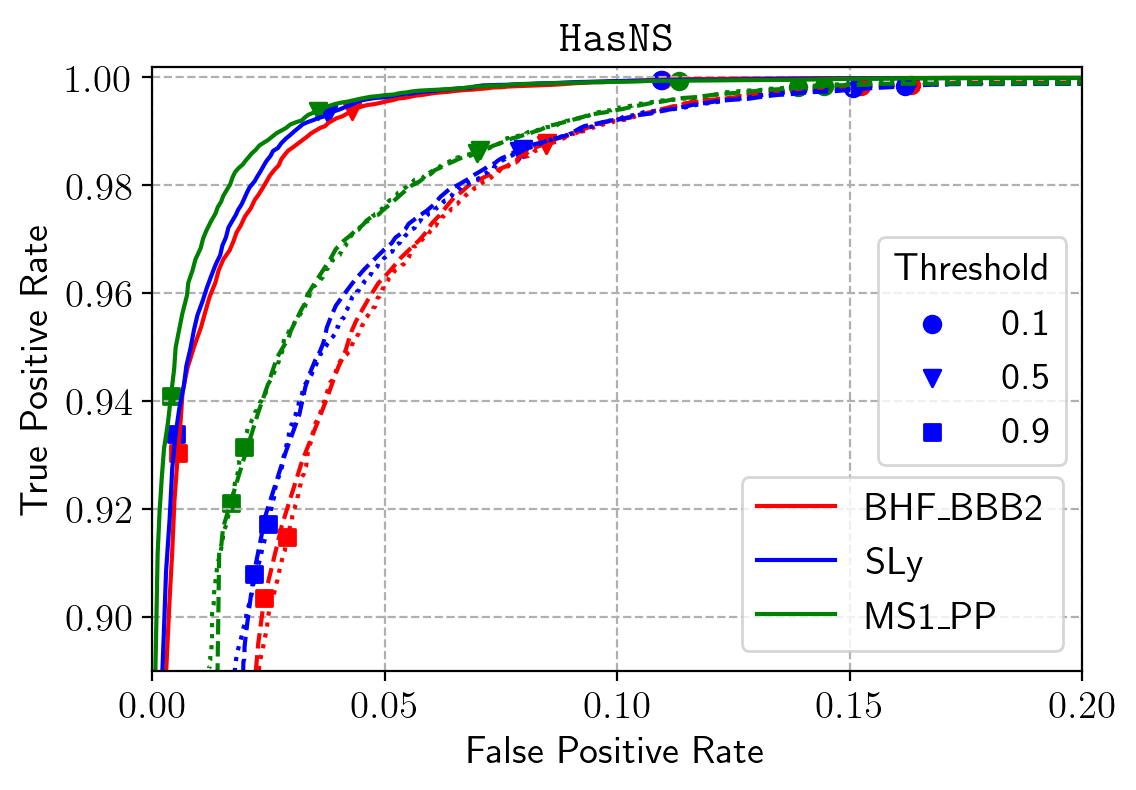}
\includegraphics[width=0.45\linewidth]{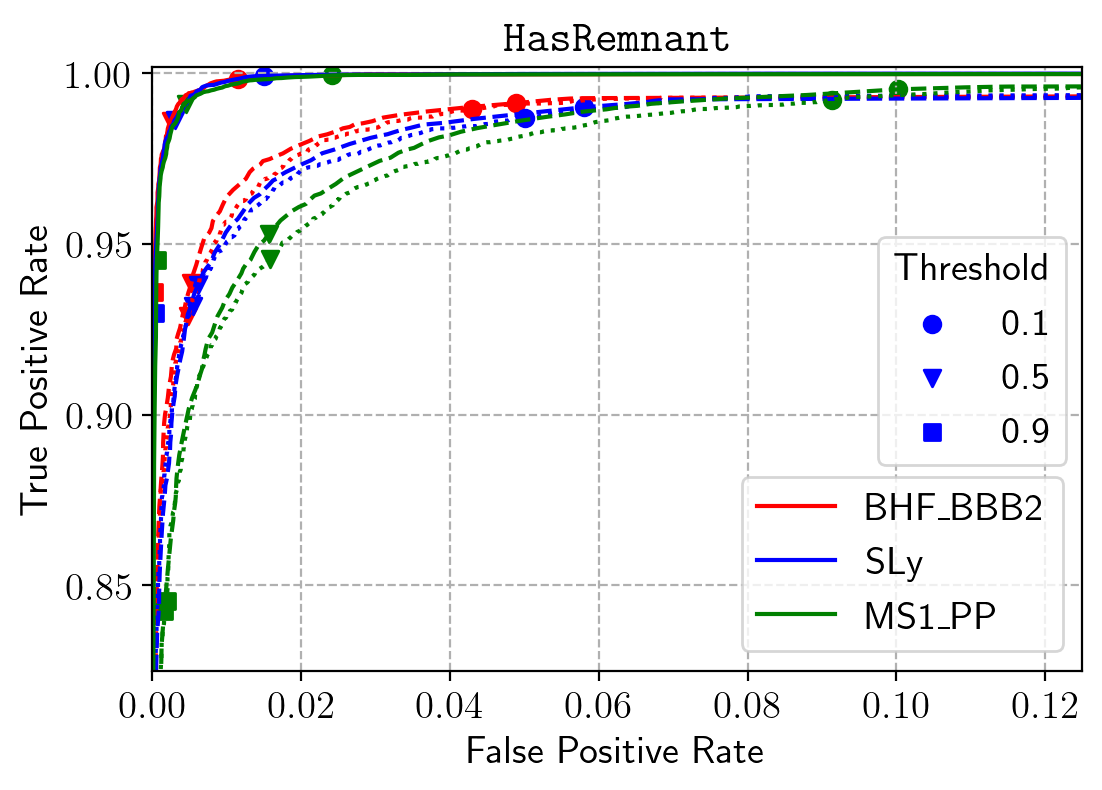}
\caption{\ac{ROC} curves derived from the \ac{O2} testing data set $D_S$ for the \ac{RF} classifier (solid lines), the \ac{KNN} implementation presented in this paper (dashed lines), and the \ac{KNN} approach currently used by the \ac{LVK} Collaborations (dotted lines). The curves for {\tt BHF\_BBB2}, {\tt MS1\_PP}, and {\tt SLy} are colored red, green, and blue, respectively. The circle, triangle, and square marks represent score thresholds of 0.1, 0.5, and 0.9, respectively.}
\label{fig:rocO2_all}
\end{figure*}

To compare the existing \ac{LVK} implementation of \ac{KNN}~\cite{Chatterjee:2019avs} with our multi-label \ac{KNN} and \ac{RF} schemes, we show the \ac{ROC} curves for the three methods in Fig.~\ref{fig:rocO2_all}. To provide a fair comparison, we trained the LVK's \ac{KNN} classifier for two cases (\hasns and \hasrem) separately, resulting in a binary label method. In this instance, we use the hyperparameters that are currently configured in the low-latency implementation: $K = 2n + 1 = 11$ neighbors (where $n$ is the number of features), and neighbor weighting by the inverse of distance. Both classifiers were trained using the same 70\% of the \ac{O2} dataset (the $D_R$ subset)\footnote{The current LVK implementation trains the \ac{KNN} algorithm on the entire dataset $D$. This leaves no events to test. Therefore, to ensure a fair comparison, we use a 70\%-30\% split.}. The ROC curve below was generated using the remaining 30\% of the dataset (the $D_S$ subset). The multi-label RF method achieves a high \ac{TPR} across the entire range of \ac{FPR} for both \hasns\ and \hasrem. Both multi-label and binary-label \ac{KNN} methods perform similarly. These \ac{ROC} curve comparisons were done in the same way as in Figures~\ref{fig:rocO2_KNN} and~\ref{fig:rocO2_RF} before defining Bayesian probabilities, as they were generated from the testing set.

\bibliography{classification.bib}

\begin{thebibliography}{73}%
\makeatletter
\providecommand \@ifxundefined [1]{%
 \@ifx{#1\undefined}
}%
\providecommand \@ifnum [1]{%
 \ifnum #1\expandafter \@firstoftwo
 \else \expandafter \@secondoftwo
 \fi
}%
\providecommand \@ifx [1]{%
 \ifx #1\expandafter \@firstoftwo
 \else \expandafter \@secondoftwo
 \fi
}%
\providecommand \natexlab [1]{#1}%
\providecommand \enquote  [1]{``#1''}%
\providecommand \bibnamefont  [1]{#1}%
\providecommand \bibfnamefont [1]{#1}%
\providecommand \citenamefont [1]{#1}%
\providecommand \href@noop [0]{\@secondoftwo}%
\providecommand \href [0]{\begingroup \@sanitize@url \@href}%
\providecommand \@href[1]{\@@startlink{#1}\@@href}%
\providecommand \@@href[1]{\endgroup#1\@@endlink}%
\providecommand \@sanitize@url [0]{\catcode `\\12\catcode `\$12\catcode
  `\&12\catcode `\#12\catcode `\^12\catcode `\_12\catcode `\%12\relax}%
\providecommand \@@startlink[1]{}%
\providecommand \@@endlink[0]{}%
\providecommand \url  [0]{\begingroup\@sanitize@url \@url }%
\providecommand \@url [1]{\endgroup\@href {#1}{\urlprefix }}%
\providecommand \urlprefix  [0]{URL }%
\providecommand \Eprint [0]{\href }%
\providecommand \doibase [0]{https://doi.org/}%
\providecommand \selectlanguage [0]{\@gobble}%
\providecommand \bibinfo  [0]{\@secondoftwo}%
\providecommand \bibfield  [0]{\@secondoftwo}%
\providecommand \translation [1]{[#1]}%
\providecommand \BibitemOpen [0]{}%
\providecommand \bibitemStop [0]{}%
\providecommand \bibitemNoStop [0]{.\EOS\space}%
\providecommand \EOS [0]{\spacefactor3000\relax}%
\providecommand \BibitemShut  [1]{\csname bibitem#1\endcsname}%
\let\auto@bib@innerbib\@empty
\bibitem [{\citenamefont {Abbott}\ \emph {et~al.}(2016)\citenamefont {Abbott}
  \emph {et~al.}}]{LIGOScientific:2016aoc}%
  \BibitemOpen
  \bibfield  {author} {\bibinfo {author} {\bibfnamefont {B.~P.}\ \bibnamefont
  {Abbott}} \emph {et~al.} (\bibinfo {collaboration} {LIGO Scientific,
  Virgo}),\ }\bibfield  {title} {\bibinfo {title} {{Observation of
  Gravitational Waves from a Binary Black Hole Merger}},\ }\href
  {https://doi.org/10.1103/PhysRevLett.116.061102} {\bibfield  {journal}
  {\bibinfo  {journal} {Phys. Rev. Lett.}\ }\textbf {\bibinfo {volume} {116}},\
  \bibinfo {pages} {061102} (\bibinfo {year} {2016})},\ \Eprint
  {https://arxiv.org/abs/1602.03837} {arXiv:1602.03837 [gr-qc]} \BibitemShut
  {NoStop}%
\bibitem [{\citenamefont {Abbott}\ \emph
  {et~al.}(2017{\natexlab{a}})\citenamefont {Abbott} \emph
  {et~al.}}]{LIGOScientific:2017vwq}%
  \BibitemOpen
  \bibfield  {author} {\bibinfo {author} {\bibfnamefont {B.~P.}\ \bibnamefont
  {Abbott}} \emph {et~al.} (\bibinfo {collaboration} {LIGO Scientific,
  Virgo}),\ }\bibfield  {title} {\bibinfo {title} {{GW170817: Observation of
  Gravitational Waves from a Binary Neutron Star Inspiral}},\ }\href
  {https://doi.org/10.1103/PhysRevLett.119.161101} {\bibfield  {journal}
  {\bibinfo  {journal} {Phys. Rev. Lett.}\ }\textbf {\bibinfo {volume} {119}},\
  \bibinfo {pages} {161101} (\bibinfo {year} {2017}{\natexlab{a}})},\ \Eprint
  {https://arxiv.org/abs/1710.05832} {arXiv:1710.05832 [gr-qc]} \BibitemShut
  {NoStop}%
\bibitem [{\citenamefont {Abbott}\ \emph
  {et~al.}(2018{\natexlab{a}})\citenamefont {Abbott} \emph
  {et~al.}}]{LIGOScientific:2018cki}%
  \BibitemOpen
  \bibfield  {author} {\bibinfo {author} {\bibfnamefont {B.~P.}\ \bibnamefont
  {Abbott}} \emph {et~al.} (\bibinfo {collaboration} {LIGO Scientific,
  Virgo}),\ }\bibfield  {title} {\bibinfo {title} {{GW170817: Measurements of
  neutron star radii and equation of state}},\ }\href
  {https://doi.org/10.1103/PhysRevLett.121.161101} {\bibfield  {journal}
  {\bibinfo  {journal} {Phys. Rev. Lett.}\ }\textbf {\bibinfo {volume} {121}},\
  \bibinfo {pages} {161101} (\bibinfo {year} {2018}{\natexlab{a}})},\ \Eprint
  {https://arxiv.org/abs/1805.11581} {arXiv:1805.11581 [gr-qc]} \BibitemShut
  {NoStop}%
\bibitem [{\citenamefont {Abbott}\ \emph
  {et~al.}(2021{\natexlab{a}})\citenamefont {Abbott} \emph
  {et~al.}}]{LIGOScientific:2021djp}%
  \BibitemOpen
  \bibfield  {author} {\bibinfo {author} {\bibfnamefont {R.}~\bibnamefont
  {Abbott}} \emph {et~al.} (\bibinfo {collaboration} {LIGO Scientific, VIRGO,
  KAGRA}),\ }\bibfield  {title} {\bibinfo {title} {{GWTC-3: Compact Binary
  Coalescences Observed by LIGO and Virgo During the Second Part of the Third
  Observing Run}},\ }\href@noop {} {\  (\bibinfo {year}
  {2021}{\natexlab{a}})},\ \Eprint {https://arxiv.org/abs/2111.03606}
  {arXiv:2111.03606 [gr-qc]} \BibitemShut {NoStop}%
\bibitem [{\citenamefont {Ruiz}\ \emph {et~al.}(2021)\citenamefont {Ruiz},
  \citenamefont {Shapiro},\ and\ \citenamefont {Tsokaros}}]{Ruiz:2021gsv}%
  \BibitemOpen
  \bibfield  {author} {\bibinfo {author} {\bibfnamefont {M.}~\bibnamefont
  {Ruiz}}, \bibinfo {author} {\bibfnamefont {S.~L.}\ \bibnamefont {Shapiro}},\
  and\ \bibinfo {author} {\bibfnamefont {A.}~\bibnamefont {Tsokaros}},\
  }\bibfield  {title} {\bibinfo {title} {{Multimessenger Binary Mergers
  Containing Neutron Stars: Gravitational Waves, Jets, and $\gamma$-Ray
  Bursts}},\ }\href {https://doi.org/10.3389/fspas.2021.656907} {\bibfield
  {journal} {\bibinfo  {journal} {Front. Astron. Space Sci.}\ }\textbf
  {\bibinfo {volume} {8}},\ \bibinfo {pages} {39} (\bibinfo {year} {2021})},\
  \Eprint {https://arxiv.org/abs/2102.03366} {arXiv:2102.03366 [astro-ph.HE]}
  \BibitemShut {NoStop}%
\bibitem [{\citenamefont {Baiotti}\ and\ \citenamefont
  {Rezzolla}(2017)}]{Baiotti:2016qnr}%
  \BibitemOpen
  \bibfield  {author} {\bibinfo {author} {\bibfnamefont {L.}~\bibnamefont
  {Baiotti}}\ and\ \bibinfo {author} {\bibfnamefont {L.}~\bibnamefont
  {Rezzolla}},\ }\bibfield  {title} {\bibinfo {title} {{Binary neutron star
  mergers: a review of Einstein\textquoteright{}s richest laboratory}},\ }\href
  {https://doi.org/10.1088/1361-6633/aa67bb} {\bibfield  {journal} {\bibinfo
  {journal} {Rept. Prog. Phys.}\ }\textbf {\bibinfo {volume} {80}},\ \bibinfo
  {pages} {096901} (\bibinfo {year} {2017})},\ \Eprint
  {https://arxiv.org/abs/1607.03540} {arXiv:1607.03540 [gr-qc]} \BibitemShut
  {NoStop}%
\bibitem [{\citenamefont {Lasky}(2015)}]{Lasky:2015uia}%
  \BibitemOpen
  \bibfield  {author} {\bibinfo {author} {\bibfnamefont {P.~D.}\ \bibnamefont
  {Lasky}},\ }\bibfield  {title} {\bibinfo {title} {{Gravitational Waves from
  Neutron Stars: A Review}},\ }\href {https://doi.org/10.1017/pasa.2015.35}
  {\bibfield  {journal} {\bibinfo  {journal} {Publ. Astron. Soc. Austral.}\
  }\textbf {\bibinfo {volume} {32}},\ \bibinfo {pages} {e034} (\bibinfo {year}
  {2015})},\ \Eprint {https://arxiv.org/abs/1508.06643} {arXiv:1508.06643
  [astro-ph.HE]} \BibitemShut {NoStop}%
\bibitem [{\citenamefont {Murase}\ and\ \citenamefont
  {Bartos}(2019)}]{Murase:2019tjj}%
  \BibitemOpen
  \bibfield  {author} {\bibinfo {author} {\bibfnamefont {K.}~\bibnamefont
  {Murase}}\ and\ \bibinfo {author} {\bibfnamefont {I.}~\bibnamefont
  {Bartos}},\ }\bibfield  {title} {\bibinfo {title} {{High-Energy
  Multimessenger Transient Astrophysics}},\ }\href
  {https://doi.org/10.1146/annurev-nucl-101918-023510} {\bibfield  {journal}
  {\bibinfo  {journal} {Ann. Rev. Nucl. Part. Sci.}\ }\textbf {\bibinfo
  {volume} {69}},\ \bibinfo {pages} {477} (\bibinfo {year} {2019})},\ \Eprint
  {https://arxiv.org/abs/1907.12506} {arXiv:1907.12506 [astro-ph.HE]}
  \BibitemShut {NoStop}%
\bibitem [{\citenamefont {Ciolfi}(2018)}]{Ciolfi:2018tal}%
  \BibitemOpen
  \bibfield  {author} {\bibinfo {author} {\bibfnamefont {R.}~\bibnamefont
  {Ciolfi}},\ }\bibfield  {title} {\bibinfo {title} {{Short gamma-ray burst
  central engines}},\ }\href {https://doi.org/10.1142/S021827181842004X}
  {\bibfield  {journal} {\bibinfo  {journal} {Int. J. Mod. Phys. D}\ }\textbf
  {\bibinfo {volume} {27}},\ \bibinfo {pages} {1842004} (\bibinfo {year}
  {2018})},\ \Eprint {https://arxiv.org/abs/1804.03684} {arXiv:1804.03684
  [astro-ph.HE]} \BibitemShut {NoStop}%
\bibitem [{\citenamefont {Schmidt}(2020)}]{Schmidt:2020ekt}%
  \BibitemOpen
  \bibfield  {author} {\bibinfo {author} {\bibfnamefont {P.}~\bibnamefont
  {Schmidt}},\ }\bibfield  {title} {\bibinfo {title} {{Gravitational Waves From
  Binary Black Hole Mergers: Modeling and Observations}},\ }\href
  {https://doi.org/10.3389/fspas.2020.00028} {\bibfield  {journal} {\bibinfo
  {journal} {Front. Astron. Space Sci.}\ }\textbf {\bibinfo {volume} {7}},\
  \bibinfo {pages} {28} (\bibinfo {year} {2020})}\BibitemShut {NoStop}%
\bibitem [{\citenamefont {Nitz}\ \emph {et~al.}(2023)\citenamefont {Nitz},
  \citenamefont {Kumar}, \citenamefont {Wang}, \citenamefont {Kastha},
  \citenamefont {Wu}, \citenamefont {Sch\"afer}, \citenamefont {Dhurkunde},\
  and\ \citenamefont {Capano}}]{Nitz:2021zwj}%
  \BibitemOpen
  \bibfield  {author} {\bibinfo {author} {\bibfnamefont {A.~H.}\ \bibnamefont
  {Nitz}}, \bibinfo {author} {\bibfnamefont {S.}~\bibnamefont {Kumar}},
  \bibinfo {author} {\bibfnamefont {Y.-F.}\ \bibnamefont {Wang}}, \bibinfo
  {author} {\bibfnamefont {S.}~\bibnamefont {Kastha}}, \bibinfo {author}
  {\bibfnamefont {S.}~\bibnamefont {Wu}}, \bibinfo {author} {\bibfnamefont
  {M.}~\bibnamefont {Sch\"afer}}, \bibinfo {author} {\bibfnamefont
  {R.}~\bibnamefont {Dhurkunde}},\ and\ \bibinfo {author} {\bibfnamefont
  {C.~D.}\ \bibnamefont {Capano}},\ }\bibfield  {title} {\bibinfo {title}
  {{4-OGC: Catalog of Gravitational Waves from Compact Binary Mergers}},\
  }\href {https://doi.org/10.3847/1538-4357/aca591} {\bibfield  {journal}
  {\bibinfo  {journal} {Astrophys. J.}\ }\textbf {\bibinfo {volume} {946}},\
  \bibinfo {pages} {59} (\bibinfo {year} {2023})},\ \Eprint
  {https://arxiv.org/abs/2112.06878} {arXiv:2112.06878 [astro-ph.HE]}
  \BibitemShut {NoStop}%
\bibitem [{\citenamefont {Barack}\ \emph {et~al.}(2019)\citenamefont {Barack}
  \emph {et~al.}}]{Barack:2018yly}%
  \BibitemOpen
  \bibfield  {author} {\bibinfo {author} {\bibfnamefont {L.}~\bibnamefont
  {Barack}} \emph {et~al.},\ }\bibfield  {title} {\bibinfo {title} {{Black
  holes, gravitational waves and fundamental physics: a roadmap}},\ }\href
  {https://doi.org/10.1088/1361-6382/ab0587} {\bibfield  {journal} {\bibinfo
  {journal} {Class. Quant. Grav.}\ }\textbf {\bibinfo {volume} {36}},\ \bibinfo
  {pages} {143001} (\bibinfo {year} {2019})},\ \Eprint
  {https://arxiv.org/abs/1806.05195} {arXiv:1806.05195 [gr-qc]} \BibitemShut
  {NoStop}%
\bibitem [{\citenamefont {Abbott}\ \emph
  {et~al.}(2021{\natexlab{b}})\citenamefont {Abbott} \emph
  {et~al.}}]{LIGOScientific:2021sio}%
  \BibitemOpen
  \bibfield  {author} {\bibinfo {author} {\bibfnamefont {R.}~\bibnamefont
  {Abbott}} \emph {et~al.} (\bibinfo {collaboration} {LIGO Scientific, VIRGO,
  KAGRA}),\ }\bibfield  {title} {\bibinfo {title} {{Tests of General Relativity
  with GWTC-3}},\ }\href@noop {} {\  (\bibinfo {year} {2021}{\natexlab{b}})},\
  \Eprint {https://arxiv.org/abs/2112.06861} {arXiv:2112.06861 [gr-qc]}
  \BibitemShut {NoStop}%
\bibitem [{\citenamefont {Berti}\ \emph
  {et~al.}(2018{\natexlab{a}})\citenamefont {Berti}, \citenamefont {Yagi},\
  and\ \citenamefont {Yunes}}]{Berti:2018cxi}%
  \BibitemOpen
  \bibfield  {author} {\bibinfo {author} {\bibfnamefont {E.}~\bibnamefont
  {Berti}}, \bibinfo {author} {\bibfnamefont {K.}~\bibnamefont {Yagi}},\ and\
  \bibinfo {author} {\bibfnamefont {N.}~\bibnamefont {Yunes}},\ }\bibfield
  {title} {\bibinfo {title} {{Extreme Gravity Tests with Gravitational Waves
  from Compact Binary Coalescences: (I) Inspiral-Merger}},\ }\href
  {https://doi.org/10.1007/s10714-018-2362-8} {\bibfield  {journal} {\bibinfo
  {journal} {Gen. Rel. Grav.}\ }\textbf {\bibinfo {volume} {50}},\ \bibinfo
  {pages} {46} (\bibinfo {year} {2018}{\natexlab{a}})},\ \Eprint
  {https://arxiv.org/abs/1801.03208} {arXiv:1801.03208 [gr-qc]} \BibitemShut
  {NoStop}%
\bibitem [{\citenamefont {Berti}\ \emph
  {et~al.}(2018{\natexlab{b}})\citenamefont {Berti}, \citenamefont {Yagi},
  \citenamefont {Yang},\ and\ \citenamefont {Yunes}}]{Berti:2018vdi}%
  \BibitemOpen
  \bibfield  {author} {\bibinfo {author} {\bibfnamefont {E.}~\bibnamefont
  {Berti}}, \bibinfo {author} {\bibfnamefont {K.}~\bibnamefont {Yagi}},
  \bibinfo {author} {\bibfnamefont {H.}~\bibnamefont {Yang}},\ and\ \bibinfo
  {author} {\bibfnamefont {N.}~\bibnamefont {Yunes}},\ }\bibfield  {title}
  {\bibinfo {title} {{Extreme Gravity Tests with Gravitational Waves from
  Compact Binary Coalescences: (II) Ringdown}},\ }\href
  {https://doi.org/10.1007/s10714-018-2372-6} {\bibfield  {journal} {\bibinfo
  {journal} {Gen. Rel. Grav.}\ }\textbf {\bibinfo {volume} {50}},\ \bibinfo
  {pages} {49} (\bibinfo {year} {2018}{\natexlab{b}})},\ \Eprint
  {https://arxiv.org/abs/1801.03587} {arXiv:1801.03587 [gr-qc]} \BibitemShut
  {NoStop}%
\bibitem [{\citenamefont {Isi}\ \emph {et~al.}(2019)\citenamefont {Isi},
  \citenamefont {Giesler}, \citenamefont {Farr}, \citenamefont {Scheel},\ and\
  \citenamefont {Teukolsky}}]{Isi:2019aib}%
  \BibitemOpen
  \bibfield  {author} {\bibinfo {author} {\bibfnamefont {M.}~\bibnamefont
  {Isi}}, \bibinfo {author} {\bibfnamefont {M.}~\bibnamefont {Giesler}},
  \bibinfo {author} {\bibfnamefont {W.~M.}\ \bibnamefont {Farr}}, \bibinfo
  {author} {\bibfnamefont {M.~A.}\ \bibnamefont {Scheel}},\ and\ \bibinfo
  {author} {\bibfnamefont {S.~A.}\ \bibnamefont {Teukolsky}},\ }\bibfield
  {title} {\bibinfo {title} {{Testing the no-hair theorem with GW150914}},\
  }\href {https://doi.org/10.1103/PhysRevLett.123.111102} {\bibfield  {journal}
  {\bibinfo  {journal} {Phys. Rev. Lett.}\ }\textbf {\bibinfo {volume} {123}},\
  \bibinfo {pages} {111102} (\bibinfo {year} {2019})},\ \Eprint
  {https://arxiv.org/abs/1905.00869} {arXiv:1905.00869 [gr-qc]} \BibitemShut
  {NoStop}%
\bibitem [{\citenamefont {Seoane}\ \emph {et~al.}(2023)\citenamefont {Seoane}
  \emph {et~al.}}]{LISA:2022yao}%
  \BibitemOpen
  \bibfield  {author} {\bibinfo {author} {\bibfnamefont {P.~A.}\ \bibnamefont
  {Seoane}} \emph {et~al.} (\bibinfo {collaboration} {LISA}),\ }\bibfield
  {title} {\bibinfo {title} {{Astrophysics with the Laser Interferometer Space
  Antenna}},\ }\href {https://doi.org/10.1007/s41114-022-00041-y} {\bibfield
  {journal} {\bibinfo  {journal} {Living Rev. Rel.}\ }\textbf {\bibinfo
  {volume} {26}},\ \bibinfo {pages} {2} (\bibinfo {year} {2023})},\ \Eprint
  {https://arxiv.org/abs/2203.06016} {arXiv:2203.06016 [gr-qc]} \BibitemShut
  {NoStop}%
\bibitem [{\citenamefont {Barausse}\ \emph {et~al.}(2020)\citenamefont
  {Barausse} \emph {et~al.}}]{Barausse:2020rsu}%
  \BibitemOpen
  \bibfield  {author} {\bibinfo {author} {\bibfnamefont {E.}~\bibnamefont
  {Barausse}} \emph {et~al.},\ }\bibfield  {title} {\bibinfo {title}
  {{Prospects for Fundamental Physics with LISA}},\ }\href
  {https://doi.org/10.1007/s10714-020-02691-1} {\bibfield  {journal} {\bibinfo
  {journal} {Gen. Rel. Grav.}\ }\textbf {\bibinfo {volume} {52}},\ \bibinfo
  {pages} {81} (\bibinfo {year} {2020})},\ \Eprint
  {https://arxiv.org/abs/2001.09793} {arXiv:2001.09793 [gr-qc]} \BibitemShut
  {NoStop}%
\bibitem [{\citenamefont {Pi\'orkowska-Kurpas}\ and\ \citenamefont
  {Biesiada}(2022)}]{Piorkowska-Kurpas:2022idw}%
  \BibitemOpen
  \bibfield  {author} {\bibinfo {author} {\bibfnamefont {A.}~\bibnamefont
  {Pi\'orkowska-Kurpas}}\ and\ \bibinfo {author} {\bibfnamefont
  {M.}~\bibnamefont {Biesiada}},\ }\bibfield  {title} {\bibinfo {title}
  {{Testing Quantum Gravity in the Multi-Messenger Astronomy Era}},\ }\href
  {https://doi.org/10.3390/universe8060321} {\bibfield  {journal} {\bibinfo
  {journal} {Universe}\ }\textbf {\bibinfo {volume} {8}},\ \bibinfo {pages}
  {321} (\bibinfo {year} {2022})}\BibitemShut {NoStop}%
\bibitem [{\citenamefont {Abbott}\ \emph
  {et~al.}(2021{\natexlab{c}})\citenamefont {Abbott} \emph
  {et~al.}}]{LIGOScientific:2021psn}%
  \BibitemOpen
  \bibfield  {author} {\bibinfo {author} {\bibfnamefont {R.}~\bibnamefont
  {Abbott}} \emph {et~al.} (\bibinfo {collaboration} {LIGO Scientific, VIRGO,
  KAGRA}),\ }\bibfield  {title} {\bibinfo {title} {{The population of merging
  compact binaries inferred using gravitational waves through GWTC-3}},\
  }\href@noop {} {\  (\bibinfo {year} {2021}{\natexlab{c}})},\ \Eprint
  {https://arxiv.org/abs/2111.03634} {arXiv:2111.03634 [astro-ph.HE]}
  \BibitemShut {NoStop}%
\bibitem [{\citenamefont {Abbott}\ \emph
  {et~al.}(2021{\natexlab{d}})\citenamefont {Abbott} \emph
  {et~al.}}]{LIGOScientific:2021aug}%
  \BibitemOpen
  \bibfield  {author} {\bibinfo {author} {\bibfnamefont {R.}~\bibnamefont
  {Abbott}} \emph {et~al.} (\bibinfo {collaboration} {LIGO Scientific, VIRGO,
  KAGRA}),\ }\bibfield  {title} {\bibinfo {title} {{Constraints on the cosmic
  expansion history from GWTC-3}},\ }\href@noop {} {\  (\bibinfo {year}
  {2021}{\natexlab{d}})},\ \Eprint {https://arxiv.org/abs/2111.03604}
  {arXiv:2111.03604 [astro-ph.CO]} \BibitemShut {NoStop}%
\bibitem [{\citenamefont {Aasi}\ \emph {et~al.}(2015)\citenamefont {Aasi} \emph
  {et~al.}}]{LIGOScientific:2014pky}%
  \BibitemOpen
  \bibfield  {author} {\bibinfo {author} {\bibfnamefont {J.}~\bibnamefont
  {Aasi}} \emph {et~al.} (\bibinfo {collaboration} {LIGO Scientific}),\
  }\bibfield  {title} {\bibinfo {title} {{Advanced LIGO}},\ }\href
  {https://doi.org/10.1088/0264-9381/32/7/074001} {\bibfield  {journal}
  {\bibinfo  {journal} {Class. Quant. Grav.}\ }\textbf {\bibinfo {volume}
  {32}},\ \bibinfo {pages} {074001} (\bibinfo {year} {2015})},\ \Eprint
  {https://arxiv.org/abs/1411.4547} {arXiv:1411.4547 [gr-qc]} \BibitemShut
  {NoStop}%
\bibitem [{\citenamefont {Acernese}\ \emph {et~al.}(2015)\citenamefont
  {Acernese} \emph {et~al.}}]{VIRGO:2014yos}%
  \BibitemOpen
  \bibfield  {author} {\bibinfo {author} {\bibfnamefont {F.}~\bibnamefont
  {Acernese}} \emph {et~al.} (\bibinfo {collaboration} {VIRGO}),\ }\bibfield
  {title} {\bibinfo {title} {{Advanced Virgo: a second-generation
  interferometric gravitational wave detector}},\ }\href
  {https://doi.org/10.1088/0264-9381/32/2/024001} {\bibfield  {journal}
  {\bibinfo  {journal} {Class. Quant. Grav.}\ }\textbf {\bibinfo {volume}
  {32}},\ \bibinfo {pages} {024001} (\bibinfo {year} {2015})},\ \Eprint
  {https://arxiv.org/abs/1408.3978} {arXiv:1408.3978 [gr-qc]} \BibitemShut
  {NoStop}%
\bibitem [{\citenamefont {Abbott}\ \emph
  {et~al.}(2018{\natexlab{b}})\citenamefont {Abbott} \emph
  {et~al.}}]{KAGRA:2013rdx}%
  \BibitemOpen
  \bibfield  {author} {\bibinfo {author} {\bibfnamefont {B.~P.}\ \bibnamefont
  {Abbott}} \emph {et~al.} (\bibinfo {collaboration} {KAGRA, LIGO Scientific,
  Virgo, VIRGO}),\ }\bibfield  {title} {\bibinfo {title} {{Prospects for
  observing and localizing gravitational-wave transients with Advanced LIGO,
  Advanced Virgo and KAGRA}},\ }\href
  {https://doi.org/10.1007/s41114-020-00026-9} {\bibfield  {journal} {\bibinfo
  {journal} {Living Rev. Rel.}\ }\textbf {\bibinfo {volume} {21}},\ \bibinfo
  {pages} {3} (\bibinfo {year} {2018}{\natexlab{b}})},\ \Eprint
  {https://arxiv.org/abs/1304.0670} {arXiv:1304.0670 [gr-qc]} \BibitemShut
  {NoStop}%
\bibitem [{\citenamefont {Lattimer}\ and\ \citenamefont
  {Schramm}(1974)}]{Lattimer:1974slx}%
  \BibitemOpen
  \bibfield  {author} {\bibinfo {author} {\bibfnamefont {J.~M.}\ \bibnamefont
  {Lattimer}}\ and\ \bibinfo {author} {\bibfnamefont {D.~N.}\ \bibnamefont
  {Schramm}},\ }\bibfield  {title} {\bibinfo {title} {{Black-hole-neutron-star
  collisions}},\ }\href {https://doi.org/10.1086/181612} {\bibfield  {journal}
  {\bibinfo  {journal} {Astrophys. J. Lett.}\ }\textbf {\bibinfo {volume}
  {192}},\ \bibinfo {pages} {L145} (\bibinfo {year} {1974})}\BibitemShut
  {NoStop}%
\bibitem [{\citenamefont {Li}\ and\ \citenamefont
  {Paczynski}(1998)}]{Li:1998bw}%
  \BibitemOpen
  \bibfield  {author} {\bibinfo {author} {\bibfnamefont {L.-X.}\ \bibnamefont
  {Li}}\ and\ \bibinfo {author} {\bibfnamefont {B.}~\bibnamefont {Paczynski}},\
  }\bibfield  {title} {\bibinfo {title} {{Transient events from neutron star
  mergers}},\ }\href {https://doi.org/10.1086/311680} {\bibfield  {journal}
  {\bibinfo  {journal} {Astrophys. J. Lett.}\ }\textbf {\bibinfo {volume}
  {507}},\ \bibinfo {pages} {L59} (\bibinfo {year} {1998})},\ \Eprint
  {https://arxiv.org/abs/astro-ph/9807272} {arXiv:astro-ph/9807272}
  \BibitemShut {NoStop}%
\bibitem [{\citenamefont {Korobkin}\ \emph {et~al.}(2012)\citenamefont
  {Korobkin}, \citenamefont {Rosswog}, \citenamefont {Arcones},\ and\
  \citenamefont {Winteler}}]{Korobkin:2012uy}%
  \BibitemOpen
  \bibfield  {author} {\bibinfo {author} {\bibfnamefont {O.}~\bibnamefont
  {Korobkin}}, \bibinfo {author} {\bibfnamefont {S.}~\bibnamefont {Rosswog}},
  \bibinfo {author} {\bibfnamefont {A.}~\bibnamefont {Arcones}},\ and\ \bibinfo
  {author} {\bibfnamefont {C.}~\bibnamefont {Winteler}},\ }\bibfield  {title}
  {\bibinfo {title} {{On the astrophysical robustness of neutron star merger
  r-process}},\ }\href {https://doi.org/10.1111/j.1365-2966.2012.21859.x}
  {\bibfield  {journal} {\bibinfo  {journal} {Mon. Not. Roy. Astron. Soc.}\
  }\textbf {\bibinfo {volume} {426}},\ \bibinfo {pages} {1940} (\bibinfo {year}
  {2012})},\ \Eprint {https://arxiv.org/abs/1206.2379} {arXiv:1206.2379
  [astro-ph.SR]} \BibitemShut {NoStop}%
\bibitem [{\citenamefont {Barnes}\ and\ \citenamefont
  {Kasen}(2013)}]{Barnes:2013wka}%
  \BibitemOpen
  \bibfield  {author} {\bibinfo {author} {\bibfnamefont {J.}~\bibnamefont
  {Barnes}}\ and\ \bibinfo {author} {\bibfnamefont {D.}~\bibnamefont {Kasen}},\
  }\bibfield  {title} {\bibinfo {title} {{Effect of a High Opacity on the Light
  Curves of Radioactively Powered Transients from Compact Object Mergers}},\
  }\href {https://doi.org/10.1088/0004-637X/775/1/18} {\bibfield  {journal}
  {\bibinfo  {journal} {Astrophys. J.}\ }\textbf {\bibinfo {volume} {775}},\
  \bibinfo {pages} {18} (\bibinfo {year} {2013})},\ \Eprint
  {https://arxiv.org/abs/1303.5787} {arXiv:1303.5787 [astro-ph.HE]}
  \BibitemShut {NoStop}%
\bibitem [{\citenamefont {Tanaka}\ and\ \citenamefont
  {Hotokezaka}(2013)}]{Tanaka:2013ana}%
  \BibitemOpen
  \bibfield  {author} {\bibinfo {author} {\bibfnamefont {M.}~\bibnamefont
  {Tanaka}}\ and\ \bibinfo {author} {\bibfnamefont {K.}~\bibnamefont
  {Hotokezaka}},\ }\bibfield  {title} {\bibinfo {title} {{Radiative Transfer
  Simulations of Neutron Star Merger Ejecta}},\ }\href
  {https://doi.org/10.1088/0004-637X/775/2/113} {\bibfield  {journal} {\bibinfo
   {journal} {Astrophys. J.}\ }\textbf {\bibinfo {volume} {775}},\ \bibinfo
  {pages} {113} (\bibinfo {year} {2013})},\ \Eprint
  {https://arxiv.org/abs/1306.3742} {arXiv:1306.3742 [astro-ph.HE]}
  \BibitemShut {NoStop}%
\bibitem [{\citenamefont {Kasen}\ \emph {et~al.}(2015)\citenamefont {Kasen},
  \citenamefont {Fernandez},\ and\ \citenamefont {Metzger}}]{Kasen:2014toa}%
  \BibitemOpen
  \bibfield  {author} {\bibinfo {author} {\bibfnamefont {D.}~\bibnamefont
  {Kasen}}, \bibinfo {author} {\bibfnamefont {R.}~\bibnamefont {Fernandez}},\
  and\ \bibinfo {author} {\bibfnamefont {B.}~\bibnamefont {Metzger}},\
  }\bibfield  {title} {\bibinfo {title} {{Kilonova light curves from the disc
  wind outflows of compact object mergers}},\ }\href
  {https://doi.org/10.1093/mnras/stv721} {\bibfield  {journal} {\bibinfo
  {journal} {Mon. Not. Roy. Astron. Soc.}\ }\textbf {\bibinfo {volume} {450}},\
  \bibinfo {pages} {1777} (\bibinfo {year} {2015})},\ \Eprint
  {https://arxiv.org/abs/1411.3726} {arXiv:1411.3726 [astro-ph.HE]}
  \BibitemShut {NoStop}%
\bibitem [{\citenamefont {Abbott}\ \emph
  {et~al.}(2017{\natexlab{b}})\citenamefont {Abbott} \emph
  {et~al.}}]{LIGOScientific:2017ync}%
  \BibitemOpen
  \bibfield  {author} {\bibinfo {author} {\bibfnamefont {B.~P.}\ \bibnamefont
  {Abbott}} \emph {et~al.} (\bibinfo {collaboration} {LIGO Scientific, Virgo,
  Fermi GBM, INTEGRAL, IceCube, AstroSat Cadmium Zinc Telluride Imager Team,
  IPN, Insight-Hxmt, ANTARES, Swift, AGILE Team, 1M2H Team, Dark Energy Camera
  GW-EM, DES, DLT40, GRAWITA, Fermi-LAT, ATCA, ASKAP, Las Cumbres Observatory
  Group, OzGrav, DWF (Deeper Wider Faster Program), AST3, CAASTRO, VINROUGE,
  MASTER, J-GEM, GROWTH, JAGWAR, CaltechNRAO, TTU-NRAO, NuSTAR, Pan-STARRS,
  MAXI Team, TZAC Consortium, KU, Nordic Optical Telescope, ePESSTO, GROND,
  Texas Tech University, SALT Group, TOROS, BOOTES, MWA, CALET, IKI-GW
  Follow-up, H.E.S.S., LOFAR, LWA, HAWC, Pierre Auger, ALMA, Euro VLBI Team, Pi
  of Sky, Chandra Team at McGill University, DFN, ATLAS Telescopes, High Time
  Resolution Universe Survey, RIMAS, RATIR, SKA South Africa/MeerKAT}),\
  }\bibfield  {title} {\bibinfo {title} {{Multi-messenger Observations of a
  Binary Neutron Star Merger}},\ }\href
  {https://doi.org/10.3847/2041-8213/aa91c9} {\bibfield  {journal} {\bibinfo
  {journal} {Astrophys. J. Lett.}\ }\textbf {\bibinfo {volume} {848}},\
  \bibinfo {pages} {L12} (\bibinfo {year} {2017}{\natexlab{b}})},\ \Eprint
  {https://arxiv.org/abs/1710.05833} {arXiv:1710.05833 [astro-ph.HE]}
  \BibitemShut {NoStop}%
\bibitem [{\citenamefont {Arcavi}\ \emph {et~al.}(2017)\citenamefont {Arcavi}
  \emph {et~al.}}]{Arcavi:2017xiz}%
  \BibitemOpen
  \bibfield  {author} {\bibinfo {author} {\bibfnamefont {I.}~\bibnamefont
  {Arcavi}} \emph {et~al.},\ }\bibfield  {title} {\bibinfo {title} {{Optical
  emission from a kilonova following a gravitational-wave-detected neutron-star
  merger}},\ }\href {https://doi.org/10.1038/nature24291} {\bibfield  {journal}
  {\bibinfo  {journal} {Nature}\ }\textbf {\bibinfo {volume} {551}},\ \bibinfo
  {pages} {64} (\bibinfo {year} {2017})},\ \Eprint
  {https://arxiv.org/abs/1710.05843} {arXiv:1710.05843 [astro-ph.HE]}
  \BibitemShut {NoStop}%
\bibitem [{\citenamefont {Coulter}\ \emph {et~al.}(2017)\citenamefont {Coulter}
  \emph {et~al.}}]{Coulter:2017wya}%
  \BibitemOpen
  \bibfield  {author} {\bibinfo {author} {\bibfnamefont {D.~A.}\ \bibnamefont
  {Coulter}} \emph {et~al.},\ }\bibfield  {title} {\bibinfo {title} {{Swope
  Supernova Survey 2017a (SSS17a), the Optical Counterpart to a Gravitational
  Wave Source}},\ }\href {https://doi.org/10.1126/science.aap9811} {\bibfield
  {journal} {\bibinfo  {journal} {Science}\ }\textbf {\bibinfo {volume}
  {358}},\ \bibinfo {pages} {1556} (\bibinfo {year} {2017})},\ \Eprint
  {https://arxiv.org/abs/1710.05452} {arXiv:1710.05452 [astro-ph.HE]}
  \BibitemShut {NoStop}%
\bibitem [{\citenamefont {Kasliwal}\ \emph {et~al.}(2017)\citenamefont
  {Kasliwal} \emph {et~al.}}]{Kasliwal:2017ngb}%
  \BibitemOpen
  \bibfield  {author} {\bibinfo {author} {\bibfnamefont {M.~M.}\ \bibnamefont
  {Kasliwal}} \emph {et~al.},\ }\bibfield  {title} {\bibinfo {title}
  {{Illuminating Gravitational Waves: A Concordant Picture of Photons from a
  Neutron Star Merger}},\ }\href {https://doi.org/10.1126/science.aap9455}
  {\bibfield  {journal} {\bibinfo  {journal} {Science}\ }\textbf {\bibinfo
  {volume} {358}},\ \bibinfo {pages} {1559} (\bibinfo {year} {2017})},\ \Eprint
  {https://arxiv.org/abs/1710.05436} {arXiv:1710.05436 [astro-ph.HE]}
  \BibitemShut {NoStop}%
\bibitem [{\citenamefont {Lipunov}\ \emph {et~al.}(2017)\citenamefont {Lipunov}
  \emph {et~al.}}]{Lipunov:2017dwd}%
  \BibitemOpen
  \bibfield  {author} {\bibinfo {author} {\bibfnamefont {V.~M.}\ \bibnamefont
  {Lipunov}} \emph {et~al.},\ }\bibfield  {title} {\bibinfo {title} {{MASTER
  Optical Detection of the First LIGO/Virgo Neutron Star Binary Merger
  GW170817}},\ }\href {https://doi.org/10.3847/2041-8213/aa92c0} {\bibfield
  {journal} {\bibinfo  {journal} {Astrophys. J. Lett.}\ }\textbf {\bibinfo
  {volume} {850}},\ \bibinfo {pages} {L1} (\bibinfo {year} {2017})},\ \Eprint
  {https://arxiv.org/abs/1710.05461} {arXiv:1710.05461 [astro-ph.HE]}
  \BibitemShut {NoStop}%
\bibitem [{\citenamefont {Soares-Santos}\ \emph {et~al.}(2017)\citenamefont
  {Soares-Santos} \emph {et~al.}}]{DES:2017kbs}%
  \BibitemOpen
  \bibfield  {author} {\bibinfo {author} {\bibfnamefont {M.}~\bibnamefont
  {Soares-Santos}} \emph {et~al.} (\bibinfo {collaboration} {DES, Dark Energy
  Camera GW-EM}),\ }\bibfield  {title} {\bibinfo {title} {{The Electromagnetic
  Counterpart of the Binary Neutron Star Merger LIGO/Virgo GW170817. I.
  Discovery of the Optical Counterpart Using the Dark Energy Camera}},\ }\href
  {https://doi.org/10.3847/2041-8213/aa9059} {\bibfield  {journal} {\bibinfo
  {journal} {Astrophys. J. Lett.}\ }\textbf {\bibinfo {volume} {848}},\
  \bibinfo {pages} {L16} (\bibinfo {year} {2017})},\ \Eprint
  {https://arxiv.org/abs/1710.05459} {arXiv:1710.05459 [astro-ph.HE]}
  \BibitemShut {NoStop}%
\bibitem [{\citenamefont {Tanvir}\ \emph {et~al.}(2017)\citenamefont {Tanvir}
  \emph {et~al.}}]{Tanvir:2017pws}%
  \BibitemOpen
  \bibfield  {author} {\bibinfo {author} {\bibfnamefont {N.~R.}\ \bibnamefont
  {Tanvir}} \emph {et~al.},\ }\bibfield  {title} {\bibinfo {title} {{The
  Emergence of a Lanthanide-Rich Kilonova Following the Merger of Two Neutron
  Stars}},\ }\href {https://doi.org/10.3847/2041-8213/aa90b6} {\bibfield
  {journal} {\bibinfo  {journal} {Astrophys. J. Lett.}\ }\textbf {\bibinfo
  {volume} {848}},\ \bibinfo {pages} {L27} (\bibinfo {year} {2017})},\ \Eprint
  {https://arxiv.org/abs/1710.05455} {arXiv:1710.05455 [astro-ph.HE]}
  \BibitemShut {NoStop}%
\bibitem [{\citenamefont {Sachdev}\ \emph {et~al.}(2019)\citenamefont {Sachdev}
  \emph {et~al.}}]{Sachdev:2019vvd}%
  \BibitemOpen
  \bibfield  {author} {\bibinfo {author} {\bibfnamefont {S.}~\bibnamefont
  {Sachdev}} \emph {et~al.},\ }\bibfield  {title} {\bibinfo {title} {{The
  GstLAL Search Analysis Methods for Compact Binary Mergers in Advanced LIGO's
  Second and Advanced Virgo's First Observing Runs}},\ }\href@noop {}
  {\bibfield  {journal} {\bibinfo  {journal} {{}}\ } (\bibinfo {year}
  {2019})},\ \Eprint {https://arxiv.org/abs/1901.08580} {arXiv:1901.08580
  [gr-qc]} \BibitemShut {NoStop}%
\bibitem [{\citenamefont {Messick}\ \emph {et~al.}(2017)\citenamefont
  {Messick}, \citenamefont {Blackburn}, \citenamefont {Brady}, \citenamefont
  {Brockill}, \citenamefont {Cannon}, \citenamefont {Cariou}, \citenamefont
  {Caudill}, \citenamefont {Chamberlin}, \citenamefont {Creighton},
  \citenamefont {Everett}, \citenamefont {Hanna}, \citenamefont {Keppel},
  \citenamefont {Lang}, \citenamefont {Li}, \citenamefont {Meacher},
  \citenamefont {Nielsen}, \citenamefont {Pankow}, \citenamefont {Privitera},
  \citenamefont {Qi}, \citenamefont {Sachdev}, \citenamefont {Sadeghian},
  \citenamefont {Singer}, \citenamefont {Thomas}, \citenamefont {Wade},
  \citenamefont {Wade}, \citenamefont {Weinstein},\ and\ \citenamefont
  {Wiesner}}]{PhysRevD.95.042001}%
  \BibitemOpen
  \bibfield  {author} {\bibinfo {author} {\bibfnamefont {C.}~\bibnamefont
  {Messick}}, \bibinfo {author} {\bibfnamefont {K.}~\bibnamefont {Blackburn}},
  \bibinfo {author} {\bibfnamefont {P.}~\bibnamefont {Brady}}, \bibinfo
  {author} {\bibfnamefont {P.}~\bibnamefont {Brockill}}, \bibinfo {author}
  {\bibfnamefont {K.}~\bibnamefont {Cannon}}, \bibinfo {author} {\bibfnamefont
  {R.}~\bibnamefont {Cariou}}, \bibinfo {author} {\bibfnamefont
  {S.}~\bibnamefont {Caudill}}, \bibinfo {author} {\bibfnamefont {S.~J.}\
  \bibnamefont {Chamberlin}}, \bibinfo {author} {\bibfnamefont {J.~D.~E.}\
  \bibnamefont {Creighton}}, \bibinfo {author} {\bibfnamefont {R.}~\bibnamefont
  {Everett}}, \bibinfo {author} {\bibfnamefont {C.}~\bibnamefont {Hanna}},
  \bibinfo {author} {\bibfnamefont {D.}~\bibnamefont {Keppel}}, \bibinfo
  {author} {\bibfnamefont {R.~N.}\ \bibnamefont {Lang}}, \bibinfo {author}
  {\bibfnamefont {T.~G.~F.}\ \bibnamefont {Li}}, \bibinfo {author}
  {\bibfnamefont {D.}~\bibnamefont {Meacher}}, \bibinfo {author} {\bibfnamefont
  {A.}~\bibnamefont {Nielsen}}, \bibinfo {author} {\bibfnamefont
  {C.}~\bibnamefont {Pankow}}, \bibinfo {author} {\bibfnamefont
  {S.}~\bibnamefont {Privitera}}, \bibinfo {author} {\bibfnamefont
  {H.}~\bibnamefont {Qi}}, \bibinfo {author} {\bibfnamefont {S.}~\bibnamefont
  {Sachdev}}, \bibinfo {author} {\bibfnamefont {L.}~\bibnamefont {Sadeghian}},
  \bibinfo {author} {\bibfnamefont {L.}~\bibnamefont {Singer}}, \bibinfo
  {author} {\bibfnamefont {E.~G.}\ \bibnamefont {Thomas}}, \bibinfo {author}
  {\bibfnamefont {L.}~\bibnamefont {Wade}}, \bibinfo {author} {\bibfnamefont
  {M.}~\bibnamefont {Wade}}, \bibinfo {author} {\bibfnamefont {A.}~\bibnamefont
  {Weinstein}},\ and\ \bibinfo {author} {\bibfnamefont {K.}~\bibnamefont
  {Wiesner}},\ }\bibfield  {title} {\bibinfo {title} {Analysis framework for
  the prompt discovery of compact binary mergers in gravitational-wave data},\
  }\href {https://doi.org/10.1103/PhysRevD.95.042001} {\bibfield  {journal}
  {\bibinfo  {journal} {Phys. Rev. D}\ }\textbf {\bibinfo {volume} {95}},\
  \bibinfo {pages} {042001} (\bibinfo {year} {2017})}\BibitemShut {NoStop}%
\bibitem [{\citenamefont {Sachdev}\ \emph {et~al.}(2020)\citenamefont {Sachdev}
  \emph {et~al.}}]{Sachdev:2020lfd}%
  \BibitemOpen
  \bibfield  {author} {\bibinfo {author} {\bibfnamefont {S.}~\bibnamefont
  {Sachdev}} \emph {et~al.},\ }\bibfield  {title} {\bibinfo {title} {{An
  Early-warning System for Electromagnetic Follow-up of Gravitational-wave
  Events}},\ }\href {https://doi.org/10.3847/2041-8213/abc753} {\bibfield
  {journal} {\bibinfo  {journal} {Astrophys. J. Lett.}\ }\textbf {\bibinfo
  {volume} {905}},\ \bibinfo {pages} {L25} (\bibinfo {year} {2020})},\ \Eprint
  {https://arxiv.org/abs/2008.04288} {arXiv:2008.04288 [astro-ph.HE]}
  \BibitemShut {NoStop}%
\bibitem [{\citenamefont {Nitz}\ \emph {et~al.}(2018)\citenamefont {Nitz},
  \citenamefont {Dal~Canton}, \citenamefont {Davis},\ and\ \citenamefont
  {Reyes}}]{Nitz:2018rgo}%
  \BibitemOpen
  \bibfield  {author} {\bibinfo {author} {\bibfnamefont {A.~H.}\ \bibnamefont
  {Nitz}}, \bibinfo {author} {\bibfnamefont {T.}~\bibnamefont {Dal~Canton}},
  \bibinfo {author} {\bibfnamefont {D.}~\bibnamefont {Davis}},\ and\ \bibinfo
  {author} {\bibfnamefont {S.}~\bibnamefont {Reyes}},\ }\bibfield  {title}
  {\bibinfo {title} {{Rapid detection of gravitational waves from compact
  binary mergers with PyCBC Live}},\ }\href
  {https://doi.org/10.1103/PhysRevD.98.024050} {\bibfield  {journal} {\bibinfo
  {journal} {Phys. Rev. D}\ }\textbf {\bibinfo {volume} {98}},\ \bibinfo
  {pages} {024050} (\bibinfo {year} {2018})},\ \Eprint
  {https://arxiv.org/abs/1805.11174} {arXiv:1805.11174 [gr-qc]} \BibitemShut
  {NoStop}%
\bibitem [{\citenamefont {Adams}\ \emph {et~al.}(2016)\citenamefont {Adams},
  \citenamefont {Buskulic}, \citenamefont {Germain}, \citenamefont {Guidi},
  \citenamefont {Marion}, \citenamefont {Montani}, \citenamefont {Mours},
  \citenamefont {Piergiovanni},\ and\ \citenamefont {Wang}}]{Adams:2015ulm}%
  \BibitemOpen
  \bibfield  {author} {\bibinfo {author} {\bibfnamefont {T.}~\bibnamefont
  {Adams}}, \bibinfo {author} {\bibfnamefont {D.}~\bibnamefont {Buskulic}},
  \bibinfo {author} {\bibfnamefont {V.}~\bibnamefont {Germain}}, \bibinfo
  {author} {\bibfnamefont {G.~M.}\ \bibnamefont {Guidi}}, \bibinfo {author}
  {\bibfnamefont {F.}~\bibnamefont {Marion}}, \bibinfo {author} {\bibfnamefont
  {M.}~\bibnamefont {Montani}}, \bibinfo {author} {\bibfnamefont
  {B.}~\bibnamefont {Mours}}, \bibinfo {author} {\bibfnamefont
  {F.}~\bibnamefont {Piergiovanni}},\ and\ \bibinfo {author} {\bibfnamefont
  {G.}~\bibnamefont {Wang}},\ }\bibfield  {title} {\bibinfo {title}
  {{Low-latency analysis pipeline for compact binary coalescences in the
  advanced gravitational wave detector era}},\ }\href
  {https://doi.org/10.1088/0264-9381/33/17/175012} {\bibfield  {journal}
  {\bibinfo  {journal} {Class. Quant. Grav.}\ }\textbf {\bibinfo {volume}
  {33}},\ \bibinfo {pages} {175012} (\bibinfo {year} {2016})},\ \Eprint
  {https://arxiv.org/abs/1512.02864} {arXiv:1512.02864 [gr-qc]} \BibitemShut
  {NoStop}%
\bibitem [{\citenamefont {Chu}\ \emph {et~al.}(2022)\citenamefont {Chu} \emph
  {et~al.}}]{Chu:2020pjv}%
  \BibitemOpen
  \bibfield  {author} {\bibinfo {author} {\bibfnamefont {Q.}~\bibnamefont
  {Chu}} \emph {et~al.},\ }\bibfield  {title} {\bibinfo {title} {{SPIIR online
  coherent pipeline to search for gravitational waves from compact binary
  coalescences}},\ }\href {https://doi.org/10.1103/PhysRevD.105.024023}
  {\bibfield  {journal} {\bibinfo  {journal} {Phys. Rev. D}\ }\textbf {\bibinfo
  {volume} {105}},\ \bibinfo {pages} {024023} (\bibinfo {year} {2022})},\
  \Eprint {https://arxiv.org/abs/2011.06787} {arXiv:2011.06787 [gr-qc]}
  \BibitemShut {NoStop}%
\bibitem [{\citenamefont {Klimenko}\ \emph {et~al.}(2016)\citenamefont
  {Klimenko} \emph {et~al.}}]{Klimenko:2015ypf}%
  \BibitemOpen
  \bibfield  {author} {\bibinfo {author} {\bibfnamefont {S.}~\bibnamefont
  {Klimenko}} \emph {et~al.},\ }\bibfield  {title} {\bibinfo {title} {{Method
  for detection and reconstruction of gravitational wave transients with
  networks of advanced detectors}},\ }\href
  {https://doi.org/10.1103/PhysRevD.93.042004} {\bibfield  {journal} {\bibinfo
  {journal} {Phys. Rev. D}\ }\textbf {\bibinfo {volume} {93}},\ \bibinfo
  {pages} {042004} (\bibinfo {year} {2016})},\ \Eprint
  {https://arxiv.org/abs/1511.05999} {arXiv:1511.05999 [gr-qc]} \BibitemShut
  {NoStop}%
\bibitem [{\citenamefont {Foucart}(2012)}]{Foucart:2012nc}%
  \BibitemOpen
  \bibfield  {author} {\bibinfo {author} {\bibfnamefont {F.}~\bibnamefont
  {Foucart}},\ }\bibfield  {title} {\bibinfo {title} {{Black Hole-Neutron Star
  Mergers: Disk Mass Predictions}},\ }\href
  {https://doi.org/10.1103/PhysRevD.86.124007} {\bibfield  {journal} {\bibinfo
  {journal} {Phys. Rev. D}\ }\textbf {\bibinfo {volume} {86}},\ \bibinfo
  {pages} {124007} (\bibinfo {year} {2012})},\ \Eprint
  {https://arxiv.org/abs/1207.6304} {arXiv:1207.6304 [astro-ph.HE]}
  \BibitemShut {NoStop}%
\bibitem [{\citenamefont {Foucart}\ \emph {et~al.}(2018)\citenamefont
  {Foucart}, \citenamefont {Hinderer},\ and\ \citenamefont
  {Nissanke}}]{Foucart:2018rjc}%
  \BibitemOpen
  \bibfield  {author} {\bibinfo {author} {\bibfnamefont {F.}~\bibnamefont
  {Foucart}}, \bibinfo {author} {\bibfnamefont {T.}~\bibnamefont {Hinderer}},\
  and\ \bibinfo {author} {\bibfnamefont {S.}~\bibnamefont {Nissanke}},\
  }\bibfield  {title} {\bibinfo {title} {{Remnant baryon mass in neutron
  star-black hole mergers: Predictions for binary neutron star mimickers and
  rapidly spinning black holes}},\ }\href
  {https://doi.org/10.1103/PhysRevD.98.081501} {\bibfield  {journal} {\bibinfo
  {journal} {Phys. Rev. D}\ }\textbf {\bibinfo {volume} {98}},\ \bibinfo
  {pages} {081501} (\bibinfo {year} {2018})},\ \Eprint
  {https://arxiv.org/abs/1807.00011} {arXiv:1807.00011 [astro-ph.HE]}
  \BibitemShut {NoStop}%
\bibitem [{\citenamefont {Farah}\ \emph {et~al.}(2022)\citenamefont {Farah},
  \citenamefont {Fishbach}, \citenamefont {Essick}, \citenamefont {Holz},\ and\
  \citenamefont {Galaudage}}]{Farah:2021qom}%
  \BibitemOpen
  \bibfield  {author} {\bibinfo {author} {\bibfnamefont {A.~M.}\ \bibnamefont
  {Farah}}, \bibinfo {author} {\bibfnamefont {M.}~\bibnamefont {Fishbach}},
  \bibinfo {author} {\bibfnamefont {R.}~\bibnamefont {Essick}}, \bibinfo
  {author} {\bibfnamefont {D.~E.}\ \bibnamefont {Holz}},\ and\ \bibinfo
  {author} {\bibfnamefont {S.}~\bibnamefont {Galaudage}},\ }\bibfield  {title}
  {\bibinfo {title} {{Bridging the Gap: Categorizing Gravitational-wave Events
  at the Transition between Neutron Stars and Black Holes}},\ }\href
  {https://doi.org/10.3847/1538-4357/ac5f03} {\bibfield  {journal} {\bibinfo
  {journal} {Astrophys. J.}\ }\textbf {\bibinfo {volume} {931}},\ \bibinfo
  {pages} {108} (\bibinfo {year} {2022})},\ \Eprint
  {https://arxiv.org/abs/2111.03498} {arXiv:2111.03498 [astro-ph.HE]}
  \BibitemShut {NoStop}%
\bibitem [{Use()}]{UserGuide}%
  \BibitemOpen
  \href@noop {} {}\bibinfo {howpublished}
  {\url{https://rtd.igwn.org/projects/userguide/en/v17.1/content.html}}\BibitemShut
  {NoStop}%
\bibitem [{\citenamefont {Pedregosa}\ \emph
  {et~al.}(2011{\natexlab{a}})\citenamefont {Pedregosa} \emph
  {et~al.}}]{Pedregosa:2011ork}%
  \BibitemOpen
  \bibfield  {author} {\bibinfo {author} {\bibfnamefont {F.}~\bibnamefont
  {Pedregosa}} \emph {et~al.},\ }\bibfield  {title} {\bibinfo {title}
  {{Scikit-learn: Machine Learning in Python}},\ }\href@noop {} {\bibfield
  {journal} {\bibinfo  {journal} {J. Machine Learning Res.}\ }\textbf {\bibinfo
  {volume} {12}},\ \bibinfo {pages} {2825} (\bibinfo {year}
  {2011}{\natexlab{a}})},\ \Eprint {https://arxiv.org/abs/1201.0490}
  {arXiv:1201.0490 [cs.LG]} \BibitemShut {NoStop}%
\bibitem [{\citenamefont {Chatterjee}\ \emph {et~al.}(2020)\citenamefont
  {Chatterjee}, \citenamefont {Ghosh}, \citenamefont {Brady}, \citenamefont
  {Kapadia}, \citenamefont {Miller}, \citenamefont {Nissanke},\ and\
  \citenamefont {Pannarale}}]{Chatterjee:2019avs}%
  \BibitemOpen
  \bibfield  {author} {\bibinfo {author} {\bibfnamefont {D.}~\bibnamefont
  {Chatterjee}}, \bibinfo {author} {\bibfnamefont {S.}~\bibnamefont {Ghosh}},
  \bibinfo {author} {\bibfnamefont {P.~R.}\ \bibnamefont {Brady}}, \bibinfo
  {author} {\bibfnamefont {S.~J.}\ \bibnamefont {Kapadia}}, \bibinfo {author}
  {\bibfnamefont {A.~L.}\ \bibnamefont {Miller}}, \bibinfo {author}
  {\bibfnamefont {S.}~\bibnamefont {Nissanke}},\ and\ \bibinfo {author}
  {\bibfnamefont {F.}~\bibnamefont {Pannarale}},\ }\bibfield  {title} {\bibinfo
  {title} {{A Machine Learning Based Source Property Inference for Compact
  Binary Mergers}},\ }\href {https://doi.org/10.3847/1538-4357/ab8dbe}
  {\bibfield  {journal} {\bibinfo  {journal} {Astrophys. J.}\ }\textbf
  {\bibinfo {volume} {896}},\ \bibinfo {pages} {54} (\bibinfo {year} {2020})},\
  \Eprint {https://arxiv.org/abs/1911.00116} {arXiv:1911.00116 [astro-ph.IM]}
  \BibitemShut {NoStop}%
\bibitem [{\citenamefont {Fix}\ and\ \citenamefont {Hodges}(1989)}]{Fix:1951}%
  \BibitemOpen
  \bibfield  {author} {\bibinfo {author} {\bibfnamefont {E.}~\bibnamefont
  {Fix}}\ and\ \bibinfo {author} {\bibfnamefont {J.~L.}\ \bibnamefont
  {Hodges}},\ }\bibfield  {title} {\bibinfo {title} {Discriminatory analysis.
  nonparametric discrimination: Consistency properties},\ }\href
  {http://www.jstor.org/stable/1403797} {\bibfield  {journal} {\bibinfo
  {journal} {International Statistical Review / Revue Internationale de
  Statistique}\ }\textbf {\bibinfo {volume} {57}},\ \bibinfo {pages} {238}
  (\bibinfo {year} {1989})}\BibitemShut {NoStop}%
\bibitem [{\citenamefont {Cover}\ and\ \citenamefont
  {Hart}(1967)}]{Cover:1967}%
  \BibitemOpen
  \bibfield  {author} {\bibinfo {author} {\bibfnamefont {T.}~\bibnamefont
  {Cover}}\ and\ \bibinfo {author} {\bibfnamefont {P.}~\bibnamefont {Hart}},\
  }\bibfield  {title} {\bibinfo {title} {Nearest neighbor pattern
  classification},\ }\href {https://doi.org/10.1109/TIT.1967.1053964}
  {\bibfield  {journal} {\bibinfo  {journal} {IEEE Transactions on Information
  Theory}\ }\textbf {\bibinfo {volume} {13}},\ \bibinfo {pages} {21} (\bibinfo
  {year} {1967})}\BibitemShut {NoStop}%
\bibitem [{\citenamefont {Guo}\ \emph {et~al.}(2003)\citenamefont {Guo},
  \citenamefont {Wang}, \citenamefont {Bell}, \citenamefont {Bi},\ and\
  \citenamefont {Greer}}]{Guo:2004}%
  \BibitemOpen
  \bibfield  {author} {\bibinfo {author} {\bibfnamefont {G.}~\bibnamefont
  {Guo}}, \bibinfo {author} {\bibfnamefont {H.}~\bibnamefont {Wang}}, \bibinfo
  {author} {\bibfnamefont {D.}~\bibnamefont {Bell}}, \bibinfo {author}
  {\bibfnamefont {Y.}~\bibnamefont {Bi}},\ and\ \bibinfo {author}
  {\bibfnamefont {K.}~\bibnamefont {Greer}},\ }\bibfield  {title} {\bibinfo
  {title} {Knn model-based approach in classification},\ }in\ \href
  {https://doi.org/10.1007/978-3-540-39964-3_62} {\emph {\bibinfo {booktitle}
  {{On The Move to Meaningful Internet Systems 2003: CoopIS, DOA, and
  ODBASE}}}},\ Vol.\ \bibinfo {volume} {2888}\ (\bibinfo {year} {2003})\ pp.\
  \bibinfo {pages} {986--996}\BibitemShut {NoStop}%
\bibitem [{\citenamefont {Xu}\ and\ \citenamefont {Goodacre}(2018)}]{split}%
  \BibitemOpen
  \bibfield  {author} {\bibinfo {author} {\bibfnamefont {Y.}~\bibnamefont
  {Xu}}\ and\ \bibinfo {author} {\bibfnamefont {R.}~\bibnamefont {Goodacre}},\
  }\bibfield  {title} {\bibinfo {title} {On splitting training and validation
  set: A comparative study of cross-validation, bootstrap and systematic
  sampling for estimating the generalization performance of supervised
  learning},\ }\href {https://doi.org/10.1007/s41664-018-0068-2} {\bibfield
  {journal} {\bibinfo  {journal} {Journal of Analysis and Testing}\ }\textbf
  {\bibinfo {volume} {2}} (\bibinfo {year} {2018})}\BibitemShut {NoStop}%
\bibitem [{\citenamefont {Ghosh}\ \emph {et~al.}(2021)\citenamefont {Ghosh},
  \citenamefont {Liu}, \citenamefont {Creighton}, \citenamefont {Kastaun},
  \citenamefont {Pratten},\ and\ \citenamefont {Hernandez}}]{Ghosh:2021eqv}%
  \BibitemOpen
  \bibfield  {author} {\bibinfo {author} {\bibfnamefont {S.}~\bibnamefont
  {Ghosh}}, \bibinfo {author} {\bibfnamefont {X.}~\bibnamefont {Liu}}, \bibinfo
  {author} {\bibfnamefont {J.}~\bibnamefont {Creighton}}, \bibinfo {author}
  {\bibfnamefont {W.}~\bibnamefont {Kastaun}}, \bibinfo {author} {\bibfnamefont
  {G.}~\bibnamefont {Pratten}},\ and\ \bibinfo {author} {\bibfnamefont {I.~M.}\
  \bibnamefont {Hernandez}},\ }\bibfield  {title} {\bibinfo {title} {{Rapid
  model comparison of equations of state from gravitational wave observation of
  binary neutron star coalescences}},\ }\href
  {https://doi.org/10.1103/PhysRevD.104.083003} {\bibfield  {journal} {\bibinfo
   {journal} {Phys. Rev. D}\ }\textbf {\bibinfo {volume} {104}},\ \bibinfo
  {pages} {083003} (\bibinfo {year} {2021})},\ \Eprint
  {https://arxiv.org/abs/2104.08681} {arXiv:2104.08681 [gr-qc]} \BibitemShut
  {NoStop}%
\bibitem [{\citenamefont {Chaudhary}\ \emph {et~al.}(2023)\citenamefont
  {Chaudhary} \emph {et~al.}}]{Chaudhary:2023vec}%
  \BibitemOpen
  \bibfield  {author} {\bibinfo {author} {\bibfnamefont {S.~S.}\ \bibnamefont
  {Chaudhary}} \emph {et~al.},\ }\bibfield  {title} {\bibinfo {title}
  {{Low-latency gravitational wave alert products and their performance in
  anticipation of the fourth LIGO-Virgo-KAGRA observing run}},\ }\href@noop {}
  {\  (\bibinfo {year} {2023})},\ \Eprint {https://arxiv.org/abs/2308.04545}
  {arXiv:2308.04545 [astro-ph.HE]} \BibitemShut {NoStop}%
\bibitem [{\citenamefont {Abbott}\ \emph {et~al.}(2023)\citenamefont {Abbott}
  \emph {et~al.}}]{2023ApJS..267...29A}%
  \BibitemOpen
  \bibfield  {author} {\bibinfo {author} {\bibfnamefont {B.~P.}\ \bibnamefont
  {Abbott}} \emph {et~al.} (\bibinfo {collaboration} {LIGO Scientific, Virgo,
  KAGRA}),\ }\bibfield  {title} {\bibinfo {title} {{Open Data from the Third
  Observing Run of LIGO, Virgo, KAGRA, and GEO}},\ }\href
  {https://doi.org/10.3847/1538-4365/acdc9f} {\bibfield  {journal} {\bibinfo
  {journal} {ApJS}\ }\textbf {\bibinfo {volume} {267}},\ \bibinfo {eid} {29}
  (\bibinfo {year} {2023})},\ \Eprint {https://arxiv.org/abs/2302.03676}
  {arXiv:2302.03676 [gr-qc]} \BibitemShut {NoStop}%
\bibitem [{\citenamefont {Dal~Canton}\ \emph {et~al.}(2021)\citenamefont
  {Dal~Canton}, \citenamefont {Nitz}, \citenamefont {Gadre}, \citenamefont
  {Cabourn~Davies}, \citenamefont {Villa-Ortega}, \citenamefont {Dent},
  \citenamefont {Harry},\ and\ \citenamefont {Xiao}}]{DalCanton:2020vpm}%
  \BibitemOpen
  \bibfield  {author} {\bibinfo {author} {\bibfnamefont {T.}~\bibnamefont
  {Dal~Canton}}, \bibinfo {author} {\bibfnamefont {A.~H.}\ \bibnamefont
  {Nitz}}, \bibinfo {author} {\bibfnamefont {B.}~\bibnamefont {Gadre}},
  \bibinfo {author} {\bibfnamefont {G.~S.}\ \bibnamefont {Cabourn~Davies}},
  \bibinfo {author} {\bibfnamefont {V.}~\bibnamefont {Villa-Ortega}}, \bibinfo
  {author} {\bibfnamefont {T.}~\bibnamefont {Dent}}, \bibinfo {author}
  {\bibfnamefont {I.}~\bibnamefont {Harry}},\ and\ \bibinfo {author}
  {\bibfnamefont {L.}~\bibnamefont {Xiao}},\ }\bibfield  {title} {\bibinfo
  {title} {{Real-time Search for Compact Binary Mergers in Advanced LIGO and
  Virgo's Third Observing Run Using PyCBC Live}},\ }\href
  {https://doi.org/10.3847/1538-4357/ac2f9a} {\bibfield  {journal} {\bibinfo
  {journal} {Astrophys. J.}\ }\textbf {\bibinfo {volume} {923}},\ \bibinfo
  {pages} {254} (\bibinfo {year} {2021})},\ \Eprint
  {https://arxiv.org/abs/2008.07494} {arXiv:2008.07494 [astro-ph.HE]}
  \BibitemShut {NoStop}%
\bibitem [{\citenamefont {Abbott}\ \emph
  {et~al.}(2021{\natexlab{e}})\citenamefont {Abbott} \emph
  {et~al.}}]{2021SoftX..1300658A}%
  \BibitemOpen
  \bibfield  {author} {\bibinfo {author} {\bibfnamefont {B.~P.}\ \bibnamefont
  {Abbott}} \emph {et~al.} (\bibinfo {collaboration} {LIGO Scientific, Virgo,
  KAGRA}),\ }\bibfield  {title} {\bibinfo {title} {{Open data from the first
  and second observing runs of Advanced LIGO and Advanced Virgo}},\ }\href
  {https://doi.org/10.1016/j.softx.2021.100658} {\bibfield  {journal} {\bibinfo
   {journal} {SoftwareX}\ }\textbf {\bibinfo {volume} {13}},\ \bibinfo {eid}
  {100658} (\bibinfo {year} {2021}{\natexlab{e}})},\ \Eprint
  {https://arxiv.org/abs/1912.11716} {arXiv:1912.11716 [gr-qc]} \BibitemShut
  {NoStop}%
\bibitem [{\citenamefont {{LIGO Scientific Collaboration}}\ and\ \citenamefont
  {{Virgo Collaboration}}(2017)}]{2017GCN.21509....1L}%
  \BibitemOpen
  \bibfield  {author} {\bibinfo {author} {\bibnamefont {{LIGO Scientific
  Collaboration}}}\ and\ \bibinfo {author} {\bibnamefont {{Virgo
  Collaboration}}},\ }\bibfield  {title} {\bibinfo {title} {{LIGO/Virgo
  G298048: Identification of a binary neutron star candidate coincident with
  Fermi GBM trigger 524666471/170817529}},\ }\href@noop {} {\bibfield
  {journal} {\bibinfo  {journal} {GRB Coordinates Network}\ }\textbf {\bibinfo
  {volume} {21509}},\ \bibinfo {pages} {1} (\bibinfo {year}
  {2017})}\BibitemShut {NoStop}%
\bibitem [{\citenamefont {{Ligo Scientific Collaboration}}\ and\ \citenamefont
  {{VIRGO Collaboration}}(2019{\natexlab{a}})}]{2019GCN.24168....1L}%
  \BibitemOpen
  \bibfield  {author} {\bibinfo {author} {\bibnamefont {{Ligo Scientific
  Collaboration}}}\ and\ \bibinfo {author} {\bibnamefont {{VIRGO
  Collaboration}}},\ }\bibfield  {title} {\bibinfo {title} {{LIGO/Virgo
  S190425z: Identification of a GW compact binary merger candidate.}},\
  }\href@noop {} {\bibfield  {journal} {\bibinfo  {journal} {GRB Coordinates
  Network}\ }\textbf {\bibinfo {volume} {24168}},\ \bibinfo {pages} {1}
  (\bibinfo {year} {2019}{\natexlab{a}})}\BibitemShut {NoStop}%
\bibitem [{\citenamefont {{Ligo Scientific Collaboration}}\ and\ \citenamefont
  {{VIRGO Collaboration}}(2019{\natexlab{b}})}]{2019GCN.24237....1L}%
  \BibitemOpen
  \bibfield  {author} {\bibinfo {author} {\bibnamefont {{Ligo Scientific
  Collaboration}}}\ and\ \bibinfo {author} {\bibnamefont {{VIRGO
  Collaboration}}},\ }\bibfield  {title} {\bibinfo {title} {{LIGO/Virgo
  S190426c: Identification of a GW compact binary merger candidate.}},\
  }\href@noop {} {\bibfield  {journal} {\bibinfo  {journal} {GRB Coordinates
  Network}\ }\textbf {\bibinfo {volume} {24237}},\ \bibinfo {pages} {1}
  (\bibinfo {year} {2019}{\natexlab{b}})}\BibitemShut {NoStop}%
\bibitem [{\citenamefont {{LIGO Scientific Collaboration}}\ and\ \citenamefont
  {{Virgo Collaboration}}(2019{\natexlab{a}})}]{2019GCN.25324....1L}%
  \BibitemOpen
  \bibfield  {author} {\bibinfo {author} {\bibnamefont {{LIGO Scientific
  Collaboration}}}\ and\ \bibinfo {author} {\bibnamefont {{Virgo
  Collaboration}}},\ }\bibfield  {title} {\bibinfo {title} {{LIGO/Virgo
  S190814bv: Identification of a GW compact binary merger candidate}},\
  }\href@noop {} {\bibfield  {journal} {\bibinfo  {journal} {GRB Coordinates
  Network}\ }\textbf {\bibinfo {volume} {25324}},\ \bibinfo {pages} {1}
  (\bibinfo {year} {2019}{\natexlab{a}})}\BibitemShut {NoStop}%
\bibitem [{\citenamefont {{LIGO Scientific Collaboration}}\ and\ \citenamefont
  {{Virgo Collaboration}}(2019{\natexlab{b}})}]{2019GCN.25829....1L}%
  \BibitemOpen
  \bibfield  {author} {\bibinfo {author} {\bibnamefont {{LIGO Scientific
  Collaboration}}}\ and\ \bibinfo {author} {\bibnamefont {{Virgo
  Collaboration}}},\ }\bibfield  {title} {\bibinfo {title} {{LIGO/Virgo
  S190924h: Identification of a GW compact binary merger candidate}},\
  }\href@noop {} {\bibfield  {journal} {\bibinfo  {journal} {GRB Coordinates
  Network}\ }\textbf {\bibinfo {volume} {25829}},\ \bibinfo {pages} {1}
  (\bibinfo {year} {2019}{\natexlab{b}})}\BibitemShut {NoStop}%
\bibitem [{\citenamefont {{LIGO Scientific Collaboration}}\ and\ \citenamefont
  {{Virgo Collaboration}}(2020)}]{2020GCN.26759....1L}%
  \BibitemOpen
  \bibfield  {author} {\bibinfo {author} {\bibnamefont {{LIGO Scientific
  Collaboration}}}\ and\ \bibinfo {author} {\bibnamefont {{Virgo
  Collaboration}}},\ }\bibfield  {title} {\bibinfo {title} {{LIGO/Virgo
  S200115j: Identification of a GW compact binary merger candidate}},\
  }\href@noop {} {\bibfield  {journal} {\bibinfo  {journal} {GRB Coordinates
  Network}\ }\textbf {\bibinfo {volume} {26759}},\ \bibinfo {pages} {1}
  (\bibinfo {year} {2020})}\BibitemShut {NoStop}%
\bibitem [{\citenamefont {Van~Rossum}\ and\ \citenamefont
  {Drake}(2009)}]{python3}%
  \BibitemOpen
  \bibfield  {author} {\bibinfo {author} {\bibfnamefont {G.}~\bibnamefont
  {Van~Rossum}}\ and\ \bibinfo {author} {\bibfnamefont {F.~L.}\ \bibnamefont
  {Drake}},\ }\href@noop {} {\emph {\bibinfo {title} {Python 3 Reference
  Manual}}}\ (\bibinfo  {publisher} {CreateSpace},\ \bibinfo {address} {Scotts
  Valley, CA},\ \bibinfo {year} {2009})\BibitemShut {NoStop}%
\bibitem [{pyp()}]{pypi}%
  \BibitemOpen
  \href {https://pypi.org/} {\bibinfo {title} {Python package index -
  pypi}}\BibitemShut {NoStop}%
\bibitem [{\citenamefont {Harris}\ \emph {et~al.}(2020)\citenamefont {Harris},
  \citenamefont {Millman}, \citenamefont {van~der Walt}, \citenamefont
  {Gommers}, \citenamefont {Virtanen}, \citenamefont {Cournapeau},
  \citenamefont {Wieser}, \citenamefont {Taylor}, \citenamefont {Berg},
  \citenamefont {Smith}, \citenamefont {Kern}, \citenamefont {Picus},
  \citenamefont {Hoyer}, \citenamefont {van Kerkwijk}, \citenamefont {Brett},
  \citenamefont {Haldane}, \citenamefont {del R{\'{i}}o}, \citenamefont
  {Wiebe}, \citenamefont {Peterson}, \citenamefont {G{\'{e}}rard-Marchant},
  \citenamefont {Sheppard}, \citenamefont {Reddy}, \citenamefont {Weckesser},
  \citenamefont {Abbasi}, \citenamefont {Gohlke},\ and\ \citenamefont
  {Oliphant}}]{harris2020array}%
  \BibitemOpen
  \bibfield  {author} {\bibinfo {author} {\bibfnamefont {C.~R.}\ \bibnamefont
  {Harris}}, \bibinfo {author} {\bibfnamefont {K.~J.}\ \bibnamefont {Millman}},
  \bibinfo {author} {\bibfnamefont {S.~J.}\ \bibnamefont {van~der Walt}},
  \bibinfo {author} {\bibfnamefont {R.}~\bibnamefont {Gommers}}, \bibinfo
  {author} {\bibfnamefont {P.}~\bibnamefont {Virtanen}}, \bibinfo {author}
  {\bibfnamefont {D.}~\bibnamefont {Cournapeau}}, \bibinfo {author}
  {\bibfnamefont {E.}~\bibnamefont {Wieser}}, \bibinfo {author} {\bibfnamefont
  {J.}~\bibnamefont {Taylor}}, \bibinfo {author} {\bibfnamefont
  {S.}~\bibnamefont {Berg}}, \bibinfo {author} {\bibfnamefont {N.~J.}\
  \bibnamefont {Smith}}, \bibinfo {author} {\bibfnamefont {R.}~\bibnamefont
  {Kern}}, \bibinfo {author} {\bibfnamefont {M.}~\bibnamefont {Picus}},
  \bibinfo {author} {\bibfnamefont {S.}~\bibnamefont {Hoyer}}, \bibinfo
  {author} {\bibfnamefont {M.~H.}\ \bibnamefont {van Kerkwijk}}, \bibinfo
  {author} {\bibfnamefont {M.}~\bibnamefont {Brett}}, \bibinfo {author}
  {\bibfnamefont {A.}~\bibnamefont {Haldane}}, \bibinfo {author} {\bibfnamefont
  {J.~F.}\ \bibnamefont {del R{\'{i}}o}}, \bibinfo {author} {\bibfnamefont
  {M.}~\bibnamefont {Wiebe}}, \bibinfo {author} {\bibfnamefont
  {P.}~\bibnamefont {Peterson}}, \bibinfo {author} {\bibfnamefont
  {P.}~\bibnamefont {G{\'{e}}rard-Marchant}}, \bibinfo {author} {\bibfnamefont
  {K.}~\bibnamefont {Sheppard}}, \bibinfo {author} {\bibfnamefont
  {T.}~\bibnamefont {Reddy}}, \bibinfo {author} {\bibfnamefont
  {W.}~\bibnamefont {Weckesser}}, \bibinfo {author} {\bibfnamefont
  {H.}~\bibnamefont {Abbasi}}, \bibinfo {author} {\bibfnamefont
  {C.}~\bibnamefont {Gohlke}},\ and\ \bibinfo {author} {\bibfnamefont {T.~E.}\
  \bibnamefont {Oliphant}},\ }\bibfield  {title} {\bibinfo {title} {Array
  programming with {NumPy}},\ }\href
  {https://doi.org/10.1038/s41586-020-2649-2} {\bibfield  {journal} {\bibinfo
  {journal} {Nature}\ }\textbf {\bibinfo {volume} {585}},\ \bibinfo {pages}
  {357} (\bibinfo {year} {2020})}\BibitemShut {NoStop}%
\bibitem [{\citenamefont {Virtanen}\ \emph {et~al.}(2020)\citenamefont
  {Virtanen}, \citenamefont {Gommers}, \citenamefont {Oliphant}, \citenamefont
  {Haberland}, \citenamefont {Reddy}, \citenamefont {Cournapeau}, \citenamefont
  {Burovski}, \citenamefont {Peterson}, \citenamefont {Weckesser},
  \citenamefont {Bright}, \citenamefont {{van der Walt}}, \citenamefont
  {Brett}, \citenamefont {Wilson}, \citenamefont {Millman}, \citenamefont
  {Mayorov}, \citenamefont {Nelson}, \citenamefont {Jones}, \citenamefont
  {Kern}, \citenamefont {Larson}, \citenamefont {Carey}, \citenamefont {Polat},
  \citenamefont {Feng}, \citenamefont {Moore}, \citenamefont {{VanderPlas}},
  \citenamefont {Laxalde}, \citenamefont {Perktold}, \citenamefont {Cimrman},
  \citenamefont {Henriksen}, \citenamefont {Quintero}, \citenamefont {Harris},
  \citenamefont {Archibald}, \citenamefont {Ribeiro}, \citenamefont
  {Pedregosa}, \citenamefont {{van Mulbregt}},\ and\ \citenamefont {{SciPy 1.0
  Contributors}}}]{2020SciPy-NMeth}%
  \BibitemOpen
  \bibfield  {author} {\bibinfo {author} {\bibfnamefont {P.}~\bibnamefont
  {Virtanen}}, \bibinfo {author} {\bibfnamefont {R.}~\bibnamefont {Gommers}},
  \bibinfo {author} {\bibfnamefont {T.~E.}\ \bibnamefont {Oliphant}}, \bibinfo
  {author} {\bibfnamefont {M.}~\bibnamefont {Haberland}}, \bibinfo {author}
  {\bibfnamefont {T.}~\bibnamefont {Reddy}}, \bibinfo {author} {\bibfnamefont
  {D.}~\bibnamefont {Cournapeau}}, \bibinfo {author} {\bibfnamefont
  {E.}~\bibnamefont {Burovski}}, \bibinfo {author} {\bibfnamefont
  {P.}~\bibnamefont {Peterson}}, \bibinfo {author} {\bibfnamefont
  {W.}~\bibnamefont {Weckesser}}, \bibinfo {author} {\bibfnamefont
  {J.}~\bibnamefont {Bright}}, \bibinfo {author} {\bibfnamefont {S.~J.}\
  \bibnamefont {{van der Walt}}}, \bibinfo {author} {\bibfnamefont
  {M.}~\bibnamefont {Brett}}, \bibinfo {author} {\bibfnamefont
  {J.}~\bibnamefont {Wilson}}, \bibinfo {author} {\bibfnamefont {K.~J.}\
  \bibnamefont {Millman}}, \bibinfo {author} {\bibfnamefont {N.}~\bibnamefont
  {Mayorov}}, \bibinfo {author} {\bibfnamefont {A.~R.~J.}\ \bibnamefont
  {Nelson}}, \bibinfo {author} {\bibfnamefont {E.}~\bibnamefont {Jones}},
  \bibinfo {author} {\bibfnamefont {R.}~\bibnamefont {Kern}}, \bibinfo {author}
  {\bibfnamefont {E.}~\bibnamefont {Larson}}, \bibinfo {author} {\bibfnamefont
  {C.~J.}\ \bibnamefont {Carey}}, \bibinfo {author} {\bibfnamefont
  {{\.I}.}~\bibnamefont {Polat}}, \bibinfo {author} {\bibfnamefont
  {Y.}~\bibnamefont {Feng}}, \bibinfo {author} {\bibfnamefont {E.~W.}\
  \bibnamefont {Moore}}, \bibinfo {author} {\bibfnamefont {J.}~\bibnamefont
  {{VanderPlas}}}, \bibinfo {author} {\bibfnamefont {D.}~\bibnamefont
  {Laxalde}}, \bibinfo {author} {\bibfnamefont {J.}~\bibnamefont {Perktold}},
  \bibinfo {author} {\bibfnamefont {R.}~\bibnamefont {Cimrman}}, \bibinfo
  {author} {\bibfnamefont {I.}~\bibnamefont {Henriksen}}, \bibinfo {author}
  {\bibfnamefont {E.~A.}\ \bibnamefont {Quintero}}, \bibinfo {author}
  {\bibfnamefont {C.~R.}\ \bibnamefont {Harris}}, \bibinfo {author}
  {\bibfnamefont {A.~M.}\ \bibnamefont {Archibald}}, \bibinfo {author}
  {\bibfnamefont {A.~H.}\ \bibnamefont {Ribeiro}}, \bibinfo {author}
  {\bibfnamefont {F.}~\bibnamefont {Pedregosa}}, \bibinfo {author}
  {\bibfnamefont {P.}~\bibnamefont {{van Mulbregt}}},\ and\ \bibinfo {author}
  {\bibnamefont {{SciPy 1.0 Contributors}}},\ }\bibfield  {title} {\bibinfo
  {title} {{{SciPy} 1.0: Fundamental Algorithms for Scientific Computing in
  Python}},\ }\href {https://doi.org/10.1038/s41592-019-0686-2} {\bibfield
  {journal} {\bibinfo  {journal} {Nature Methods}\ }\textbf {\bibinfo {volume}
  {17}},\ \bibinfo {pages} {261} (\bibinfo {year} {2020})}\BibitemShut
  {NoStop}%
\bibitem [{\citenamefont {pandas~development team}(2020)}]{reback2020pandas}%
  \BibitemOpen
  \bibfield  {author} {\bibinfo {author} {\bibfnamefont {T.}~\bibnamefont
  {pandas~development team}},\ }\href {https://doi.org/10.5281/zenodo.3509134}
  {\bibinfo {title} {pandas-dev/pandas: Pandas}} (\bibinfo {year}
  {2020})\BibitemShut {NoStop}%
\bibitem [{\citenamefont {{W}es
  {M}c{K}inney}(2010)}]{mckinney-proc-scipy-2010}%
  \BibitemOpen
  \bibfield  {author} {\bibinfo {author} {\bibnamefont {{W}es {M}c{K}inney}},\
  }\bibfield  {title} {\bibinfo {title} {{D}ata {S}tructures for {S}tatistical
  {C}omputing in {P}ython},\ }in\ \href
  {https://doi.org/10.25080/Majora-92bf1922-00a} {\emph {\bibinfo {booktitle}
  {{P}roceedings of the 9th {P}ython in {S}cience {C}onference}}},\ \bibinfo
  {editor} {edited by\ \bibinfo {editor} {\bibnamefont {{S}t\'efan van~der
  {W}alt}}\ and\ \bibinfo {editor} {\bibnamefont {{J}arrod {M}illman}}}\
  (\bibinfo {year} {2010})\ pp.\ \bibinfo {pages} {56 -- 61}\BibitemShut
  {NoStop}%
\bibitem [{\citenamefont {Hunter}(2007)}]{Hunter:2007ouj}%
  \BibitemOpen
  \bibfield  {author} {\bibinfo {author} {\bibfnamefont {J.~D.}\ \bibnamefont
  {Hunter}},\ }\bibfield  {title} {\bibinfo {title} {{Matplotlib: A 2D Graphics
  Environment}},\ }\href {https://doi.org/10.1109/MCSE.2007.55} {\bibfield
  {journal} {\bibinfo  {journal} {Comput. Sci. Eng.}\ }\textbf {\bibinfo
  {volume} {9}},\ \bibinfo {pages} {90} (\bibinfo {year} {2007})}\BibitemShut
  {NoStop}%
\bibitem [{\citenamefont {Pedregosa}\ \emph
  {et~al.}(2011{\natexlab{b}})\citenamefont {Pedregosa}, \citenamefont
  {Varoquaux}, \citenamefont {Gramfort}, \citenamefont {Michel}, \citenamefont
  {Thirion}, \citenamefont {Grisel}, \citenamefont {Blondel}, \citenamefont
  {Prettenhofer}, \citenamefont {Weiss}, \citenamefont {Dubourg}, \citenamefont
  {Vanderplas}, \citenamefont {Passos}, \citenamefont {Cournapeau},
  \citenamefont {Brucher}, \citenamefont {Perrot},\ and\ \citenamefont
  {Duchesnay}}]{scikit-learn}%
  \BibitemOpen
  \bibfield  {author} {\bibinfo {author} {\bibfnamefont {F.}~\bibnamefont
  {Pedregosa}}, \bibinfo {author} {\bibfnamefont {G.}~\bibnamefont
  {Varoquaux}}, \bibinfo {author} {\bibfnamefont {A.}~\bibnamefont {Gramfort}},
  \bibinfo {author} {\bibfnamefont {V.}~\bibnamefont {Michel}}, \bibinfo
  {author} {\bibfnamefont {B.}~\bibnamefont {Thirion}}, \bibinfo {author}
  {\bibfnamefont {O.}~\bibnamefont {Grisel}}, \bibinfo {author} {\bibfnamefont
  {M.}~\bibnamefont {Blondel}}, \bibinfo {author} {\bibfnamefont
  {P.}~\bibnamefont {Prettenhofer}}, \bibinfo {author} {\bibfnamefont
  {R.}~\bibnamefont {Weiss}}, \bibinfo {author} {\bibfnamefont
  {V.}~\bibnamefont {Dubourg}}, \bibinfo {author} {\bibfnamefont
  {J.}~\bibnamefont {Vanderplas}}, \bibinfo {author} {\bibfnamefont
  {A.}~\bibnamefont {Passos}}, \bibinfo {author} {\bibfnamefont
  {D.}~\bibnamefont {Cournapeau}}, \bibinfo {author} {\bibfnamefont
  {M.}~\bibnamefont {Brucher}}, \bibinfo {author} {\bibfnamefont
  {M.}~\bibnamefont {Perrot}},\ and\ \bibinfo {author} {\bibfnamefont
  {E.}~\bibnamefont {Duchesnay}},\ }\bibfield  {title} {\bibinfo {title}
  {Scikit-learn: Machine learning in {P}ython},\ }\href@noop {} {\bibfield
  {journal} {\bibinfo  {journal} {Journal of Machine Learning Research}\
  }\textbf {\bibinfo {volume} {12}},\ \bibinfo {pages} {2825} (\bibinfo {year}
  {2011}{\natexlab{b}})}\BibitemShut {NoStop}%
\end{thebibliography}%

\end{document}